\documentclass[12pt] {article}
\pdfoutput=1
\usepackage[hyphens,spaces,obeyspaces]{url}
\usepackage{setspace}

\usepackage{epsfig}
\usepackage{amssymb}
\usepackage{amsmath}
\usepackage{multirow}
\usepackage{setspace}
\usepackage{datetime}
\usepackage{amsmath}
\usepackage{algorithm}
\usepackage{algpseudocode}
\usepackage{graphicx}
\usepackage{subfig} 
\usepackage{lipsum}
\usepackage{acronym}
\usepackage{array}
\usepackage{graphicx}
\usepackage{cite}
\usepackage{graphicx}
\usepackage{epstopdf}
\usepackage{caption}
\usepackage{balance}
\usepackage{fancybox}
\usepackage{colortbl}
\usepackage{array}
\usepackage{mystyle}
\usepackage{multirow}
\usepackage[obeyspaces,hyphens,spaces]{url}
\usepackage{threeparttable}
\usepackage[english]{babel}
\usepackage{longtable}
\usepackage{listings, lstautogobble}
\usepackage{booktabs}
\usepackage{threeparttable}
\usepackage[table,xcdraw]{xcolor}

\usepackage{amsthm}
\theoremstyle{definition}
\AtBeginEnvironment{thm}{\footnotesize}
\AtBeginEnvironment{defn}{\footnotesize}
\renewcommand{\baselinestretch}{1}

\providecommand{\keywords}[1]{\textbf{\textit{Keywords---}} #1}

{\begin{list}{}%
         {\setlength{\leftmargin}{#1}}%
         \item[]%
}
{\end{list}}

\newdateformat{monthyeardate}{%
  \monthname[\THEMONTH], \THEYEAR}
\begin{document}
\title{\Large{\textbf{Comparative Study of Approximate  Multipliers}}}
\author{
Mahmoud~Masadeh$^{1}$, Osman~Hasan$^{1,2}$, and Sofi\`ene~Tahar$^{1}$ \vspace*{2em}\\
$^{1}$Department of Electrical and Computer Engineering,\\
Concordia University, Montr\'eal, Canada  \\
\{m\_masa,o\_hasan,tahar\}@ece.concordia.ca \vspace*{2em}\\
$^{2}$School of Electrical Engineering and Computer Science,\\
National University of Science and Technology, Islamabad, Pakistan \\
\\
TECHNICAL REPORT \\
\date{March 2018}
}
\maketitle

\newpage
\begin{abstract}


Approximate multipliers are widely being advocated for energy-efficient computing in applications that exhibit an inherent tolerance to inaccuracy.  However, the inclusion of accuracy as a key design parameter, besides the performance, area and power, makes the identification of the most suitable approximate multiplier quite challenging. In this paper, we identify three major decision making factors for the selection of an approximate multipliers circuit: (1) the type of approximate full adder (FA) used to construct the multiplier, (2) the architecture, i.e., array or tree, of the multiplier and (3) the placement of sub-modules of approximate and exact multipliers in the main multiplier module. Based on these factors, we explored the design space for circuit level implementations of approximate multipliers. We used circuit level implementations of some of the most widely used approximate full adders, i.e., approximate mirror adders, XOR/XNOR based approximate full adders and Inexact adder cell. These FA cells are then used to develop circuits for the approximate high order compressors as building blocks for 8x8 array and tree multipliers. We then develop various implementations of higher bit multipliers by using a combination of exact and inaccurate 8x8 multiplier cells. All these implementations have been done using the Cadence's Spectre tool with the TSMC65nm technology. The design space of these multipliers is explored based on their power, area, delay and error and the best approximate multiplier designs are identified. The report also presents the validation of our results using an image blending application. An open source library of implemented cells and multiplier circuits are available online.

\end{abstract}

\newpage
\tableofcontents
\newpage
\section{Introduction}
\label{introduction}
	The pervasive, portable, embedded and mobile nature of present age computing systems has led to an increasing demand for ultra low power consumption, small footprint, and high performance. Approximate computing \cite{AC1} is a nascent computing paradigm that allows us to achieve these objectives by compromising the arithmetic accuracy. Many systems used in domains, like multimedia and big data analysis, exhibit an inherent tolerance to a certain level of inaccuracies in computation, and thus can benefit from approximate computing.

Functional approximation \cite{M5}, in hardware, mostly deals with the design of approximate arithmetic units, such as adders and multipliers, at different abstraction levels, i.e., transistor, gate, RTL (Register Transfer Level) and application. Some notable approximate adders include \textit{speculative adders}\cite{speculative}, \textit{segmented adders}\cite{segmented}, \textit{carry select adders}\cite{CSA} and \textit{approximate full adders} \cite{AC2}. The transistor level approximation provides the highest flexibility due to the ability to tweak most of the design parameters at this level. Various approximate full adders (FA) at the transistor level have been proposed including the mirror adders \cite{Vaibhav}, the XOR/XNOR based FA \cite{XORFA} and the inexact FA \cite{InXA}. On the other hand, most of approximate multipliers have been designed at higher levels of abstraction, i.e., gate, RTL and application. 

Approximate multipliers have been mainly designed using three techniques, i) \textit{Approximation in partial products generation}: e.g., Kulkarni et al. \cite{M5} proposed an approximate 2x2 binary multiplier at the gate level by changing a single entry in the Karnaugh-map with an error rate of $1/16$. ii) \textit{Approximation in partial product tree}: e.g., Error Tolerant Multipliers (ETM) \cite{M6} divide the input operands in  two parts, i.e., the multiplication part for the MSBs and the non-multiplication part for the LSBs, and thus omitting the generation of some partial products \cite{M1}. iii) \textit{Approximation in partial products summation}: Approximate FA cells are used to form an array multiplier, e.g., in \cite{Reddy} the approximate mirror adder has been used to develop a multiplier. Similarly, Momeni et al. \cite{M2} proposed an approximate compressor for building approximate multipliers, but this multiplier is known to give a non-zero result for zero inputs. Jiang et al.\cite{Jiang} compared the characteristics of different approximate multipliers, implemented in VHDL based on the three different techniques mentioned previously. \textit{In this work, we target approximate multipliers based on approximation in partial products summation}.

In this report, we compare the accuracy and circuit characteristics of different approximate multipliers. These multipliers are designed based on three identified decisions: (1) the type of approximate FA used to construct the multiplier, (2) the architecture of the multiplier, and (3) the placement of sub-modules of approximate and exact multipliers in the target multiplier module. We were able to design  approximate multipliers, which are suitable to applications with intrinsic error resiliency. We used these designs in an image processing application and obtained promising results, thus we believe they are applicable in other domains. The rest of the report is organized as follows: The proposed methodology of designing and evaluating approximate multipliers is explained in Section \ref{methodology}. Section \ref{sec:ApproxFA} explains the design characteristics of approximate FAs and compressors. Section \ref{sec:8X8} describes different configurations of approximate sub-modules, with different architectures. Target approximate multiplies are designed and evaluated in Section \ref{sec:MainModule}. The application of image processing is given in Section \ref{sec:application}. Finally, conclusions are drawn in Section \ref{sec:conclusion}.

\section{Proposed Methodology}
\label{methodology}

The design space for approximate multipliers based on different approximate FAs and compressors is quite huge. However, it is difficult to select the most suitable design for a specific application. Figure \ref{fig:methodology} presents an overview of our proposed methodology to build different approximate multipliers and compare their design metrics to select the most suitable design. It consists of the following steps:

\begin{enumerate}
\item \textit{Building a library of elementary approximate FAs using the TSMC65nm technology in Cadence Spectre}: We use the default transistors of this technology to build 11 approximate FA designs comprising of 5 mirror FAs, 3 XOR/XNOR gate FAs and 3 inexact FAs. To the best of our knowledge, these 11 designs are the only ones that exist in the literature at the transistor level.
 
\item \textit{Characterization and early space reduction}: We perform area, power, latency and quality characterizations of different approximate FAs to filter out non-Pareto designs. 

\begin{figure}[h]
\centering
\includegraphics[width=1.0\textwidth, height=10cm]{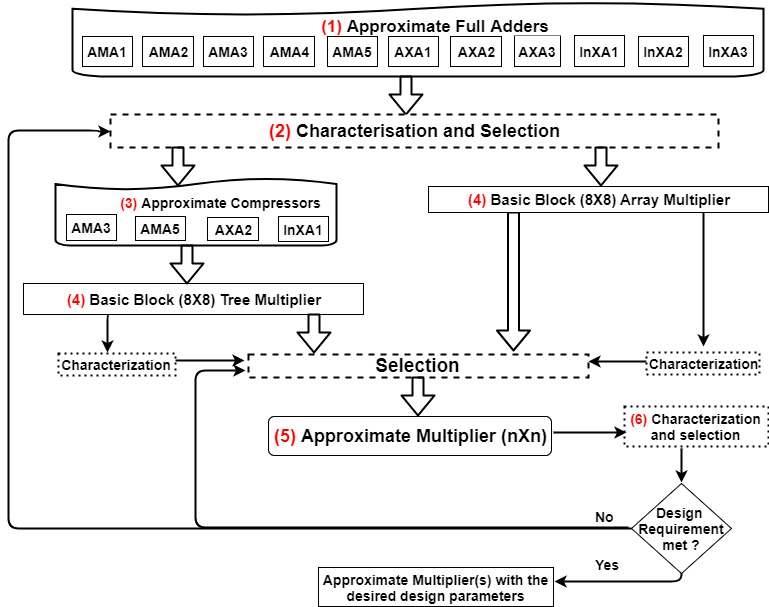}
\centering
\caption{Methodology Overview}
\label{fig:methodology}
\end{figure}

\item \textit{Building a library of approximate compressors}: We build a Cadence library of approximate compressors using the optimal approximate FA, as recommended by \cite{Vaibhav}.

\item \textit{Building approximate multipliers basic blocks}: Based on approximate FAs and compressors, we design various approximate 8x8 array and tree multipliers, respectively. These proposed designs are related to the ripple-carry array multiplier architecture, which is the most power efficient amonge conventional architectures \cite{Mrazek}.

\item \textit{Design target approximate multipliers}: Based on different configurations of 8x8 approximate multipliers, the target multiplier modules are designed and characterized.

\item \textit{Selection of design points}: Considering the required quality constrains of a specific application, a subset of power-efficient design points is selected.
\end{enumerate}

In order to evaluate the efficiency of the proposed approximate designs, \textit{power consumption} and \textit{area}, represented by the number of transistors used, are measured, and the circuit performance is measured by the maximum \textit{delay} between changing the inputs and observing the output(s). Besides these basic design metrics, \textit{accuracy} is also an important design constraint in approximate computing. There exist several \textit{error metrics} used in approximate computing to quantify errors and measure accuracy \cite{InXA}, including:
\begin{itemize}
\item Error Rate (ER): The percentage of erroneous outputs among all outputs.

\item Error Distance (ED): The arithmetic difference between the exact and approximate result.

\item Mean Error Distance (MED): The average of EDs for a set of outputs obtained by applying a set of inputs.

\item Relative Error Distance (RED): The ratio of ED to exact result. 
 
\item Mean Relative Error Distance (MRED): The average value of all possible relative error distances (RED).

\item Normalized Mean Error Distance (NMED): The normalization of mean error distance (MED) by the maximum output of the accurate design. This metric is used for comparing circuits (adders and multipliers) of different sizes. 

\end{itemize}
 
For the evaluation of the accuracy of the approximate FAs, we use the number of erroneous outputs. In the proposed methodology, we evaluate ER, MED, NMED and MRED for the proposed designs. As shown in Figure \ref{fig:methodology}, \textit{the characterization and selection} process is applied at multiple steps to different components, during the design flow. Characterization aims to find the design characteristics of the circuits including area, power consumption, performance, error metrics, and other derived metrics such as Power-Delay-Product (PDP). The design selection process for the evaluated approximate designs also depends on the application domain of the given circuit. As the design requirements vary from one application to another, our designs are unique because they can provide some degree of error in the output as well and thus this aspect also needs to be covered in the characterization and selection process.


\section{\textbf{Approximate FAs and Compressors}} \label{sec:ApproxFA}

Approximate n-bit binary adders can be designed by modifying the carry generation and propagation of the addition process by using several overlapping sub-adders to reduce latency. Some examples include speculative \cite{speculative}, segmented \cite{segmented} and carry select adders \cite{CSA}. However, these designs involve several overlapping sub-adders, which makes them unsuitable to build energy efficient circuits. Low power approximate binary adders are generally constructed by replacing the accurate FAs with approximate FAs. We consider five approximate mirror adders (AMA1, AMA2, AMA3, AMA4 and AMA5) \cite{Vaibhav}, three approximate XOR/XNOR based full adders (AXA1, AXA2 and AXA3) \cite{XORFA} and three inexact adder cells (InXA1, InXA2 and InXA3) \cite{InXA}.

Table \ref{tab:FA} shows the truth tables of the 11 considered approximate FAs, and their characteristics including Size (A), Power consumption (P), Delay (D), number of Erroneous outputs (E), which indicates the likelihood of at least one output (Cout or Sum) being wrong, and PDP. \textit{All approximate FAs are Pareto-points}, i.e., they provide less area and power consumption compared to the exact design at the cost of compromising accuracy \cite{Pareto}. Some of the FA designs have an enhanced performance (reduced delay), while other designs have degraded performance due to the internal structure and node capacitance. In \cite{Rehman}, AMA5 is considered as a \textit{wire} with zero area and zero power consumption. However, this is unrealistic as the output of AMA5 has to drive other signals. Thus, we used two buffers instead of two wires to design it.
%

\begin{table*}[th!]
  \centering
  \caption{Truth Tables of Different Approximate FAs and Comparison of their Characteristics}
\resizebox{0.99\textwidth}{!} {
    \begin{tabular}{|c|c|c|c|c|cc|cc|cc|cc|cc|cc|cc|cc|cc|cc|cc|}
    \toprule
    \multicolumn{3}{|c|}{\textbf{Inputs}} & \multicolumn{2}{c|}{\cellcolor[rgb]{ .294,  .675,  .776} \textbf{Exact FA (E)}} & \multicolumn{2}{c|}{\cellcolor[rgb]{ 0,  .69,  .314} \textbf{AMA1 (M1)}} & \multicolumn{2}{c|}{\cellcolor[rgb]{ 0,  .69,  .314} \textbf{AMA2 (M2)}} & \multicolumn{2}{c|}{\cellcolor[rgb]{ 0,  .69,  .314} \textbf{AMA3 (M3)}} & \multicolumn{2}{c|}{\cellcolor[rgb]{ 0,  .69,  .314} \textbf{AMA4 (M4)}} & \multicolumn{2}{c|}{\cellcolor[rgb]{ 0,  .69,  .314} \textbf{AMA5 (M4)}} & \multicolumn{2}{c|}{\cellcolor[rgb]{ 0,  .69,  .941} \textbf{AXA1 (X1)}} & \multicolumn{2}{c|}{\cellcolor[rgb]{ 0,  .69,  .941} \textbf{AXA2 (X2)}} & \multicolumn{2}{c|}{\cellcolor[rgb]{ 0,  .69,  .941} \textbf{AXA3 (X3)}} & \multicolumn{2}{c|}{\cellcolor[rgb]{ 1,  .753,  0} \textbf{InXA1 (In1)}} & \multicolumn{2}{c|}{\cellcolor[rgb]{ 1,  .753,  0} \textbf{InXA2 (In2)}} & \multicolumn{2}{c|}{\cellcolor[rgb]{ 1,  .753,  0} \textbf{InXA3 (In3)}} \\
    \midrule
    \textbf{A} & \textbf{B} & \textbf{Cin} & \cellcolor[rgb]{ .573,  .804,  .863} \textbf{Sum} & \cellcolor[rgb]{ .573,  .804,  .863} \textbf{Cout} & \multicolumn{1}{c|}{\cellcolor[rgb]{ .847,  .894,  .737} \textbf{Sum}} & \cellcolor[rgb]{ .847,  .894,  .737} \textbf{Cout} & \multicolumn{1}{c|}{\cellcolor[rgb]{ .847,  .894,  .737} \textbf{Sum}} & \cellcolor[rgb]{ .847,  .894,  .737} \textbf{Cout} & \multicolumn{1}{c|}{\cellcolor[rgb]{ .847,  .894,  .737} \textbf{Sum}} & \cellcolor[rgb]{ .847,  .894,  .737} \textbf{Cout} & \multicolumn{1}{c|}{\cellcolor[rgb]{ .847,  .894,  .737} \textbf{Sum}} & \cellcolor[rgb]{ .847,  .894,  .737} \textbf{Cout} & \multicolumn{1}{c|}{\cellcolor[rgb]{ .847,  .894,  .737} \textbf{Sum}} & \cellcolor[rgb]{ .847,  .894,  .737} \textbf{Cout} & \multicolumn{1}{c|}{\cellcolor[rgb]{ .773,  .851,  .945} \textbf{Sum}} & \cellcolor[rgb]{ .773,  .851,  .945} \textbf{Cout} & \multicolumn{1}{c|}{\cellcolor[rgb]{ .773,  .851,  .945} \textbf{Sum}} & \cellcolor[rgb]{ .773,  .851,  .945} \textbf{Cout} & \multicolumn{1}{c|}{\cellcolor[rgb]{ .773,  .851,  .945} \textbf{Sum}} & \cellcolor[rgb]{ .773,  .851,  .945} \textbf{Cout} & \multicolumn{1}{c|}{\cellcolor[rgb]{ .988,  .835,  .706} \textbf{Sum}} & \cellcolor[rgb]{ .988,  .835,  .706} \textbf{Cout} & \multicolumn{1}{c|}{\cellcolor[rgb]{ .988,  .835,  .706} \textbf{Sum}} & \cellcolor[rgb]{ .988,  .835,  .706} \textbf{Cout} & \multicolumn{1}{c|}{\cellcolor[rgb]{ .988,  .835,  .706} \textbf{Sum}} & \cellcolor[rgb]{ .988,  .835,  .706} \textbf{Cout} \\
    \midrule
    0  & 0 & 0  & \cellcolor[rgb]{ .851,  .851,  .851} 0 & \cellcolor[rgb]{ .851,  .851,  .851} 0 & \multicolumn{1}{c|}{0} & 0     & \multicolumn{1}{c|}{\cellcolor[rgb]{ .851,  .851,  .851} \textcolor[rgb]{ 1,  0,  0}{\textbf{1}}} & \cellcolor[rgb]{ .851,  .851,  .851} \textcolor[rgb]{ 1,  0,  0}{\textbf{0}} & \multicolumn{1}{c|}{\textcolor[rgb]{ 1,  0,  0}{\textbf{1}}} & \textcolor[rgb]{ 1,  0,  0}{\textbf{0}} & \multicolumn{1}{c|}{\cellcolor[rgb]{ .851,  .851,  .851} 0} & \cellcolor[rgb]{ .851,  .851,  .851} 0 & \multicolumn{1}{c|}{0} & 0     & \multicolumn{1}{c|}{\cellcolor[rgb]{ .851,  .851,  .851} 0} & \cellcolor[rgb]{ .851,  .851,  .851} 0 & \multicolumn{1}{c|}{\textcolor[rgb]{ 1,  0,  0}{\textbf{1}}} & \textcolor[rgb]{ 1,  0,  0}{\textbf{0}} & \multicolumn{1}{c|}{\cellcolor[rgb]{ .851,  .851,  .851} 0} & \cellcolor[rgb]{ .851,  .851,  .851} 0 & \multicolumn{1}{c|}{0} & 0     & \multicolumn{1}{c|}{\cellcolor[rgb]{ .851,  .851,  .851} 0} & \cellcolor[rgb]{ .851,  .851,  .851} 0 & \multicolumn{1}{c|}{\textcolor[rgb]{ 1,  0,  0}{\textbf{1}}} & \textcolor[rgb]{ 1,  0,  0}{\textbf{0}} \\
    \midrule
    0  & 0  & 1  & \cellcolor[rgb]{ .851,  .851,  .851} 1 & \cellcolor[rgb]{ .851,  .851,  .851} 0 & \multicolumn{1}{c|}{1} & 0     & \multicolumn{1}{c|}{\cellcolor[rgb]{ .851,  .851,  .851} 1} & \cellcolor[rgb]{ .851,  .851,  .851} 0 & \multicolumn{1}{c|}{1} & 0     & \multicolumn{1}{c|}{\cellcolor[rgb]{ .851,  .851,  .851} 1} & \cellcolor[rgb]{ .851,  .851,  .851} 0 & \multicolumn{1}{c|}{\textcolor[rgb]{ 1,  0,  0}{\textbf{0}}} & \textcolor[rgb]{ 1,  0,  0}{\textbf{0}} & \multicolumn{1}{c|}{\cellcolor[rgb]{ .851,  .851,  .851} 1} & \cellcolor[rgb]{ .851,  .851,  .851} 0 & \multicolumn{1}{c|}{1} & 0     & \multicolumn{1}{c|}{\cellcolor[rgb]{ .851,  .851,  .851} 1} & \cellcolor[rgb]{ .851,  .851,  .851} 0 & \multicolumn{1}{c|}{\textcolor[rgb]{ 1,  0,  0}{\textbf{1}}} & \textcolor[rgb]{ 1,  0,  0}{\textbf{1}} & \multicolumn{1}{c|}{\cellcolor[rgb]{ .851,  .851,  .851} 1} & \cellcolor[rgb]{ .851,  .851,  .851} 0 & \multicolumn{1}{c|}{1} & 0 \\
    \midrule
    0  & 1  & 0  & \cellcolor[rgb]{ .851,  .851,  .851} 1 & \cellcolor[rgb]{ .851,  .851,  .851} 0 & \multicolumn{1}{c|}{\textcolor[rgb]{ 1,  0,  0}{\textbf{0}}} & \textcolor[rgb]{ 1,  0,  0}{\textbf{1}} & \multicolumn{1}{c|}{\cellcolor[rgb]{ .851,  .851,  .851} 1} & \cellcolor[rgb]{ .851,  .851,  .851} 0 & \multicolumn{1}{c|}{\textcolor[rgb]{ 1,  0,  0}{\textbf{0}}} & \textcolor[rgb]{ 1,  0,  0}{\textbf{1}} & \multicolumn{1}{c|}{\cellcolor[rgb]{ .851,  .851,  .851} \textcolor[rgb]{ 1,  0,  0}{\textbf{0}}} & \cellcolor[rgb]{ .851,  .851,  .851} \textcolor[rgb]{ 1,  0,  0}{\textbf{0}} & \multicolumn{1}{c|}{1} & 0     & \multicolumn{1}{c|}{\cellcolor[rgb]{ .851,  .851,  .851} \textcolor[rgb]{ 1,  0,  0}{\textbf{0}}} & \cellcolor[rgb]{ .851,  .851,  .851} \textcolor[rgb]{ 1,  0,  0}{\textbf{1}} & \multicolumn{1}{c|}{\textcolor[rgb]{ 1,  0,  0}{\textbf{0}}} & \textcolor[rgb]{ 1,  0,  0}{\textbf{0}} & \multicolumn{1}{c|}{\cellcolor[rgb]{ .851,  .851,  .851} \textcolor[rgb]{ 1,  0,  0}{\textbf{0}}} & \cellcolor[rgb]{ .851,  .851,  .851} \textcolor[rgb]{ 1,  0,  0}{\textbf{0}} & \multicolumn{1}{c|}{1} & 0     & \multicolumn{1}{c|}{\cellcolor[rgb]{ .851,  .851,  .851} 1} & \cellcolor[rgb]{ .851,  .851,  .851} 0 & \multicolumn{1}{c|}{1} & 0 \\
    \midrule
    0  & 1  & 1  & \cellcolor[rgb]{ .851,  .851,  .851} 0 & \cellcolor[rgb]{ .851,  .851,  .851} 1 & \multicolumn{1}{c|}{0} & 1     & \multicolumn{1}{c|}{\cellcolor[rgb]{ .851,  .851,  .851} 0} & \cellcolor[rgb]{ .851,  .851,  .851} 1 & \multicolumn{1}{c|}{0} & 1     & \multicolumn{1}{c|}{\cellcolor[rgb]{ .851,  .851,  .851} \textcolor[rgb]{ 1,  0,  0}{\textbf{1}}} & \cellcolor[rgb]{ .851,  .851,  .851} \textcolor[rgb]{ 1,  0,  0}{\textbf{0}} & \multicolumn{1}{c|}{\textcolor[rgb]{ 1,  0,  0}{\textbf{1}}} & \textcolor[rgb]{ 1,  0,  0}{\textbf{0}} & \multicolumn{1}{c|}{\cellcolor[rgb]{ .851,  .851,  .851} \textcolor[rgb]{ 1,  0,  0}{\textbf{1}}} & \cellcolor[rgb]{ .851,  .851,  .851} \textcolor[rgb]{ 1,  0,  0}{\textbf{0}} & \multicolumn{1}{c|}{0} & 1     & \multicolumn{1}{c|}{\cellcolor[rgb]{ .851,  .851,  .851} 0} & \cellcolor[rgb]{ .851,  .851,  .851} 1 & \multicolumn{1}{c|}{0} & 1     & \multicolumn{1}{c|}{\cellcolor[rgb]{ .851,  .851,  .851} \textcolor[rgb]{ 1,  0,  0}{\textbf{1}}} & \cellcolor[rgb]{ .851,  .851,  .851} \textcolor[rgb]{ 1,  0,  0}{\textbf{1}} & \multicolumn{1}{c|}{0} & 1 \\
    \midrule
    1  & 0  & 0  & \cellcolor[rgb]{ .851,  .851,  .851} 1 & \cellcolor[rgb]{ .851,  .851,  .851} 0 & \multicolumn{1}{c|}{\textcolor[rgb]{ 1,  0,  0}{\textbf{0}}} & \textcolor[rgb]{ 1,  0,  0}{\textbf{0}} & \multicolumn{1}{c|}{\cellcolor[rgb]{ .851,  .851,  .851} 1} & \cellcolor[rgb]{ .851,  .851,  .851} 0 & \multicolumn{1}{c|}{1} & 0     & \multicolumn{1}{c|}{\cellcolor[rgb]{ .851,  .851,  .851} \textcolor[rgb]{ 1,  0,  0}{\textbf{0}}} & \cellcolor[rgb]{ .851,  .851,  .851} \textcolor[rgb]{ 1,  0,  0}{\textbf{1}} & \multicolumn{1}{c|}{\textcolor[rgb]{ 1,  0,  0}{\textbf{0}}} & \textcolor[rgb]{ 1,  0,  0}{\textbf{1}} & \multicolumn{1}{c|}{\cellcolor[rgb]{ .851,  .851,  .851} \textcolor[rgb]{ 1,  0,  0}{\textbf{0}}} & \cellcolor[rgb]{ .851,  .851,  .851} \textcolor[rgb]{ 1,  0,  0}{\textbf{1}} & \multicolumn{1}{c|}{\textcolor[rgb]{ 1,  0,  0}{\textbf{0}}} & \textcolor[rgb]{ 1,  0,  0}{\textbf{0}} & \multicolumn{1}{c|}{\cellcolor[rgb]{ .851,  .851,  .851} \textcolor[rgb]{ 1,  0,  0}{\textbf{0}}} & \cellcolor[rgb]{ .851,  .851,  .851} \textcolor[rgb]{ 1,  0,  0}{\textbf{0}} & \multicolumn{1}{c|}{1} & 0     & \multicolumn{1}{c|}{\cellcolor[rgb]{ .851,  .851,  .851} 1} & \cellcolor[rgb]{ .851,  .851,  .851} 0 & \multicolumn{1}{c|}{1} & 0 \\
    \midrule
    1  & 0  & 1  & \cellcolor[rgb]{ .851,  .851,  .851} 0 & \cellcolor[rgb]{ .851,  .851,  .851} 1 & \multicolumn{1}{c|}{0} & 1     & \multicolumn{1}{c|}{\cellcolor[rgb]{ .851,  .851,  .851} 0} & \cellcolor[rgb]{ .851,  .851,  .851} 1 & \multicolumn{1}{c|}{0} & 1     & \multicolumn{1}{c|}{\cellcolor[rgb]{ .851,  .851,  .851} 0} & \cellcolor[rgb]{ .851,  .851,  .851} 1 & \multicolumn{1}{c|}{0} & 1     & \multicolumn{1}{c|}{\cellcolor[rgb]{ .851,  .851,  .851} \textcolor[rgb]{ 1,  0,  0}{\textbf{1}}} & \cellcolor[rgb]{ .851,  .851,  .851} \textcolor[rgb]{ 1,  0,  0}{\textbf{0}} & \multicolumn{1}{c|}{0} & 1     & \multicolumn{1}{c|}{\cellcolor[rgb]{ .851,  .851,  .851} 0} & \cellcolor[rgb]{ .851,  .851,  .851} 1 & \multicolumn{1}{c|}{0} & 1     & \multicolumn{1}{c|}{\cellcolor[rgb]{ .851,  .851,  .851} \textcolor[rgb]{ 1,  0,  0}{\textbf{1}}} & \cellcolor[rgb]{ .851,  .851,  .851} \textcolor[rgb]{ 1,  0,  0}{\textbf{1}} & \multicolumn{1}{c|}{0} & 1 \\
    \midrule
    1  & 1  & 0  & \cellcolor[rgb]{ .851,  .851,  .851} 0 & \cellcolor[rgb]{ .851,  .851,  .851} 1 & \multicolumn{1}{c|}{0} & 1     & \multicolumn{1}{c|}{\cellcolor[rgb]{ .851,  .851,  .851} 0} & \cellcolor[rgb]{ .851,  .851,  .851} 1 & \multicolumn{1}{c|}{0} & 1     & \multicolumn{1}{c|}{\cellcolor[rgb]{ .851,  .851,  .851} 0} & \cellcolor[rgb]{ .851,  .851,  .851} 1 & \multicolumn{1}{c|}{\textcolor[rgb]{ 1,  0,  0}{\textbf{1}}} & \textcolor[rgb]{ 1,  0,  0}{\textbf{1}} & \multicolumn{1}{c|}{\cellcolor[rgb]{ .851,  .851,  .851} 0} & \cellcolor[rgb]{ .851,  .851,  .851} 1 & \multicolumn{1}{c|}{\textcolor[rgb]{ 1,  0,  0}{\textbf{1}}} & \textcolor[rgb]{ 1,  0,  0}{\textbf{1}} & \multicolumn{1}{c|}{\cellcolor[rgb]{ .851,  .851,  .851} 0} & \cellcolor[rgb]{ .851,  .851,  .851} 1 & \multicolumn{1}{c|}{\textcolor[rgb]{ 1,  0,  0}{\textbf{0}}} & \textcolor[rgb]{ 1,  0,  0}{\textbf{0}} & \multicolumn{1}{c|}{\cellcolor[rgb]{ .851,  .851,  .851} 0} & \cellcolor[rgb]{ .851,  .851,  .851} 1 & \multicolumn{1}{c|}{0} & 1 \\
    \midrule
    1  & 1  & 1  & \cellcolor[rgb]{ .851,  .851,  .851} 1 & \cellcolor[rgb]{ .851,  .851,  .851} 1 & \multicolumn{1}{c|}{1} & 1     & \multicolumn{1}{c|}{\cellcolor[rgb]{ .851,  .851,  .851} \textcolor[rgb]{ 1,  0,  0}{\textbf{0}}} & \cellcolor[rgb]{ .851,  .851,  .851} \textcolor[rgb]{ 1,  0,  0}{\textbf{1}} & \multicolumn{1}{c|}{\textcolor[rgb]{ 1,  0,  0}{\textbf{0}}} & \textcolor[rgb]{ 1,  0,  0}{\textbf{1}} & \multicolumn{1}{c|}{\cellcolor[rgb]{ .851,  .851,  .851} 1} & \cellcolor[rgb]{ .851,  .851,  .851} 1 & \multicolumn{1}{c|}{1} & 1     & \multicolumn{1}{c|}{\cellcolor[rgb]{ .851,  .851,  .851} 1} & \cellcolor[rgb]{ .851,  .851,  .851} 1 & \multicolumn{1}{c|}{1} & 1     & \multicolumn{1}{c|}{\cellcolor[rgb]{ .851,  .851,  .851} 1} & \cellcolor[rgb]{ .851,  .851,  .851} 1 & \multicolumn{1}{c|}{1} & 1     & \multicolumn{1}{c|}{\cellcolor[rgb]{ .851,  .851,  .851} 1} & \cellcolor[rgb]{ .851,  .851,  .851} 1 & \multicolumn{1}{c|}{\textcolor[rgb]{ 1,  0,  0}{\textbf{0}}} & \textcolor[rgb]{ 1,  0,  0}{\textbf{1}} \\
    \midrule
    \rowcolor[rgb]{ .902,  .722,  .718} \multicolumn{3}{|c|}{\textbf{Size}} & \multicolumn{2}{c|}{\cellcolor[rgb]{ 1,  1,  1} \textbf{28}} & \multicolumn{2}{c|}{\cellcolor[rgb]{ .769,  .843,  .608} \textbf{20}} & \multicolumn{2}{c|}{\cellcolor[rgb]{ .769,  .843,  .608} \textbf{14}} & \multicolumn{2}{c|}{\cellcolor[rgb]{ .769,  .843,  .608} \textbf{11}} & \multicolumn{2}{c|}{\cellcolor[rgb]{ .769,  .843,  .608} \textbf{15}} & \multicolumn{2}{c|}{\cellcolor[rgb]{ .769,  .843,  .608} \textbf{8}} & \multicolumn{2}{c|}{\cellcolor[rgb]{ .773,  .851,  .945} \textbf{8}} & \multicolumn{2}{c|}{\cellcolor[rgb]{ .773,  .851,  .945} \textbf{6}} & \multicolumn{2}{c|}{\cellcolor[rgb]{ .773,  .851,  .945} \textbf{8}} & \multicolumn{2}{c|}{\cellcolor[rgb]{ .988,  .835,  .706} \textbf{6}} & \multicolumn{2}{c|}{\cellcolor[rgb]{ .988,  .835,  .706} \textbf{8}} & \multicolumn{2}{c|}{\cellcolor[rgb]{ .988,  .835,  .706} \textbf{6}} \\
    \midrule
    \rowcolor[rgb]{ .902,  .722,  .718} \multicolumn{3}{|c|}{\textbf{Power (nw)}} & \multicolumn{2}{c|}{\cellcolor[rgb]{ 1,  1,  1} \textbf{763.3}} & \multicolumn{2}{c|}{\cellcolor[rgb]{ .573,  .816,  .314} \textbf{612}} & \multicolumn{2}{c|}{\cellcolor[rgb]{ .573,  .816,  .314} \textbf{561.1}} & \multicolumn{2}{c|}{\cellcolor[rgb]{ .573,  .816,  .314} \textbf{558.1}} & \multicolumn{2}{c|}{\cellcolor[rgb]{ .573,  .816,  .314} \textbf{587.1}} & \multicolumn{2}{c|}{\cellcolor[rgb]{ .573,  .816,  .314} \textbf{412.1}} & \multicolumn{2}{c|}{\cellcolor[rgb]{ .553,  .706,  .886} \textbf{676.2}} & \multicolumn{2}{c|}{\cellcolor[rgb]{ .553,  .706,  .886} \textbf{358.7}} & \multicolumn{2}{c|}{\cellcolor[rgb]{ .553,  .706,  .886} \textbf{396.5}} & \multicolumn{2}{c|}{\cellcolor[rgb]{ .992,  .914,  .851} \textbf{410}} & \multicolumn{2}{c|}{\cellcolor[rgb]{ .992,  .914,  .851} \textbf{355.1}} & \multicolumn{2}{c|}{\cellcolor[rgb]{ .992,  .914,  .851} \textbf{648}} \\
    \midrule
    \rowcolor[rgb]{ .902,  .722,  .718} \multicolumn{3}{|c|}{\textbf{Delay (ps)}} & \multicolumn{2}{c|}{\cellcolor[rgb]{ 1,  1,  1} \textbf{244}} & \multicolumn{2}{c|}{\cellcolor[rgb]{ .847,  .894,  .737} \textbf{195}} & \multicolumn{2}{c|}{\cellcolor[rgb]{ .847,  .894,  .737} \textbf{366}} & \multicolumn{2}{c|}{\cellcolor[rgb]{ .847,  .894,  .737} \textbf{360}} & \multicolumn{2}{c|}{\cellcolor[rgb]{ .847,  .894,  .737} \textbf{196}} & \multicolumn{2}{c|}{\cellcolor[rgb]{ .847,  .894,  .737} \textbf{150}} & \multicolumn{2}{c|}{\cellcolor[rgb]{ .773,  .851,  .945} \textbf{1155}} & \multicolumn{2}{c|}{\cellcolor[rgb]{ .773,  .851,  .945} \textbf{838}} & \multicolumn{2}{c|}{\cellcolor[rgb]{ .773,  .851,  .945} \textbf{1467}} & \multicolumn{2}{c|}{\cellcolor[rgb]{ .988,  .835,  .706} \textbf{740}} & \multicolumn{2}{c|}{\cellcolor[rgb]{ .988,  .835,  .706} \textbf{832}} & \multicolumn{2}{c|}{\cellcolor[rgb]{ .988,  .835,  .706} \textbf{767}} \\
    \midrule
    \rowcolor[rgb]{ .902,  .722,  .718} \multicolumn{3}{|c|}{\textbf{\# of Error Cases}} & \multicolumn{2}{c|}{\cellcolor[rgb]{ 1,  1,  1} \textbf{0}} & \multicolumn{2}{c|}{\cellcolor[rgb]{ .573,  .816,  .314} \textbf{2}} & \multicolumn{2}{c|}{\cellcolor[rgb]{ .573,  .816,  .314} \textbf{2}} & \multicolumn{2}{c|}{\cellcolor[rgb]{ .573,  .816,  .314} \textbf{3}} & \multicolumn{2}{c|}{\cellcolor[rgb]{ .573,  .816,  .314} \textbf{3}} & \multicolumn{2}{c|}{\cellcolor[rgb]{ .573,  .816,  .314} \textbf{4}} & \multicolumn{2}{c|}{\cellcolor[rgb]{ .553,  .706,  .886} \textbf{4}} & \multicolumn{2}{c|}{\cellcolor[rgb]{ .553,  .706,  .886} \textbf{4}} & \multicolumn{2}{c|}{\cellcolor[rgb]{ .553,  .706,  .886} \textbf{2}} & \multicolumn{2}{c|}{\cellcolor[rgb]{ .992,  .914,  .851} \textbf{2}} & \multicolumn{2}{c|}{\cellcolor[rgb]{ .992,  .914,  .851} \textbf{2}} & \multicolumn{2}{c|}{\cellcolor[rgb]{ .992,  .914,  .851} \textbf{2}} \\
    \midrule
    \rowcolor[rgb]{ .902,  .722,  .718} \multicolumn{3}{|c|}{\textbf{PDP (fJ)}} & \multicolumn{2}{c|}{\cellcolor[rgb]{ 1,  1,  1} \textbf{186.25}} & \multicolumn{2}{c|}{\cellcolor[rgb]{ .847,  .894,  .737} \textbf{119.34}} & \multicolumn{2}{c|}{\cellcolor[rgb]{ .847,  .894,  .737} \textbf{205.36}} & \multicolumn{2}{c|}{\cellcolor[rgb]{ .847,  .894,  .737} \textbf{200.92}} & \multicolumn{2}{c|}{\cellcolor[rgb]{ .847,  .894,  .737} \textbf{115.07}} & \multicolumn{2}{c|}{\cellcolor[rgb]{ .847,  .894,  .737} \textbf{61.82}} & \multicolumn{2}{c|}{\cellcolor[rgb]{ .773,  .851,  .945} \textbf{781}} & \multicolumn{2}{c|}{\cellcolor[rgb]{ .773,  .851,  .945} \textbf{300.59}} & \multicolumn{2}{c|}{\cellcolor[rgb]{ .773,  .851,  .945} \textbf{582}} & \multicolumn{2}{c|}{\cellcolor[rgb]{ .988,  .835,  .706} \textbf{303.4}} & \multicolumn{2}{c|}{\cellcolor[rgb]{ .988,  .835,  .706} \textbf{295.44}} & \multicolumn{2}{c|}{\cellcolor[rgb]{ .988,  .835,  .706} \textbf{753.5}} \\
    \bottomrule
    \end{tabular} }
  \label{tab:FA}
\end{table*}

\begin{figure}[!t]
\centering
\includegraphics[width=1.0\textwidth, height=5cm]{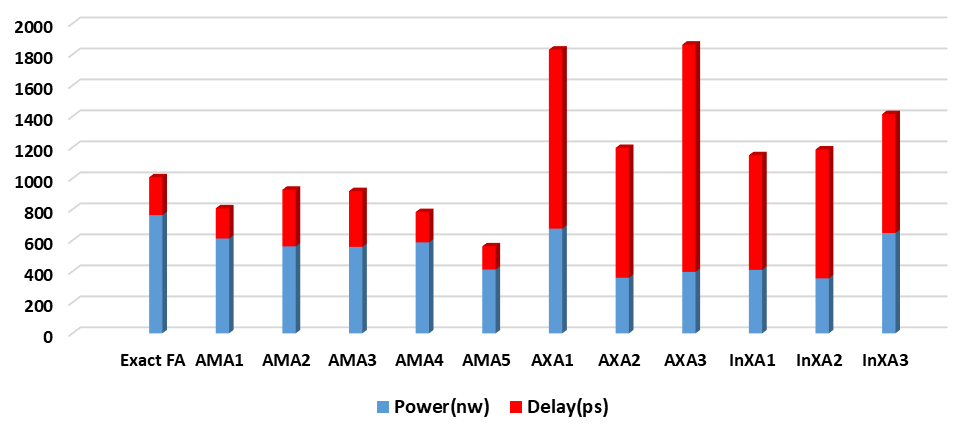}
\caption{Power Consumption and Delay of Approximate FAs}
\label{fig:Power_Delay_FA}
\end{figure}


\begin{figure}[!t]
\centering
\includegraphics[width=1.0\textwidth, height=5cm]{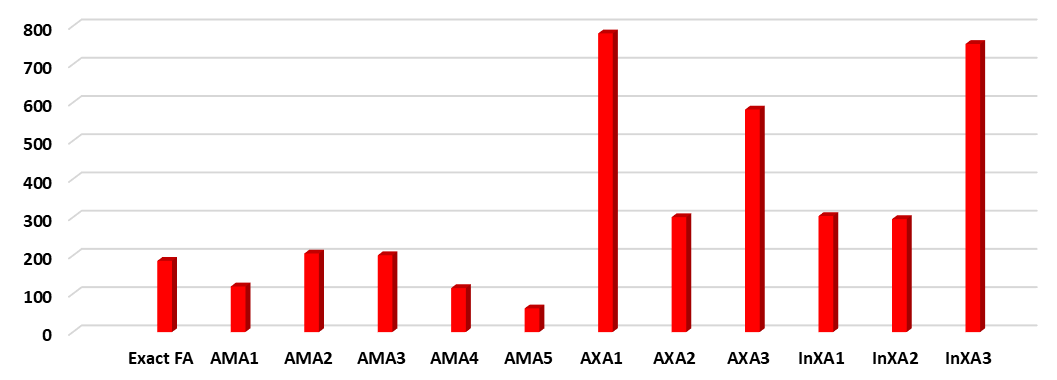}
\caption{Power-Delay-Product of Approximate FAs}
\label{fig:PDP_FA}
\end{figure}

\begin{figure}[!t]
\centering
\includegraphics[width=1.0\textwidth, height=5cm]{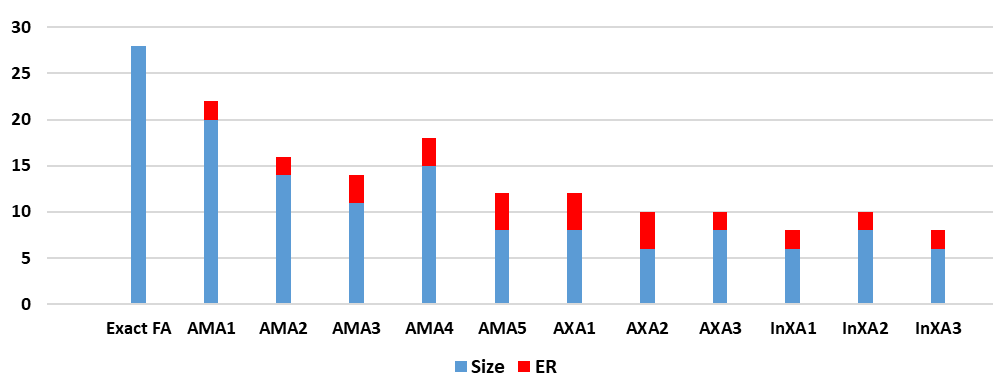}
\caption{Size and Number of erroneous outputs (ER) of Approximate FAs}
\label{fig:Area_ER_FA}
\end{figure}



Figure \ref{fig:Power_Delay_FA} shows the power consumption and delay of individual approximate FAs, where all FAs exhibit a reduced power consumption. But, only the mirror adder based FAs have a reduced delay due to their internal structure. InXA2 and AXA2 have the minimal power consumption with 53\% reduction compared to the exact mirror adder (MA). Also, InXA1 and AXA3 have close-to-minimal power consumption. Since AMA5 is composed of only two buffers, it has the lowest delay while AXA3 has the highest delay due to the threshold voltage drop of the pass transistors. AMA1 and AMA4 both have a close-to-minimal delay. PDP which is a figure of merit correlated with the energy efficiency of a digital design, is shown in Figure \ref{fig:PDP_FA} for the FAs. Mirror adder based designs have a low PDP values. AMA5 and AXA1 exhibit the lowest and highest PDP, respectively. 

Figure \ref{fig:Area_ER_FA} shows the number of transistors for each FA, as well the number of erronous outputs. AXA2, InXA1 and InXA3 consist of 6 transistors each, and thus have a 78.6\% area reduction compared to the exact MA. AMA5, AXA1, AXA3 and InXA2 all have 8 transistors. AMA5, AXA1 and AXA2 have 4 erroneous outputs. AMA3 and AMA4 have 3 erroneous outputs, and the remaining 6 designs have 2 erroneous outputs. Our results are consistent with the findings reported in \cite{Vaibhav}\cite{XORFA}\cite{InXA}.


Assuming that the characteristics of approximate FAs are linearly applied to approximate arithmetic circuits (adders and multipliers), there is no single approximate FA, which is superior in all aspects. Therefore, we propose to use a \textit{fitness function} to evaluate the designs based on its design metrics.
\begin{equation}
Fitness = \textit{C1}*A + \textit{C2}*P +  \textit{C3}*D + \textit{C4}*E + \textit{C5}*PDP 
\label{eq:fitness}
\end{equation} where \textit{C1}, \textit{C2}, \textit{C3}, \textit{C4} and \textit{C5} are application-dependent design coefficients within the range [0,1] which provide weights to specific design metrics for a specific application, e.g., E equals zero for the exact designs where approximation is not allowed, and P is small for low power designs depending of application error-resiliency. The fitness of the approximate circuit depends on the application resiliency and input data distribution. \textit{A minimal fitness value is preferred since the goal is to minimize A, P, D and E}. For the remainder of this work, we use all 11 Pareto-design approximate FAs as elementary building cells to construct approximate array multipliers. 

\begin{figure}[!t]
\centering
\includegraphics[width=8cm , height=6cm]{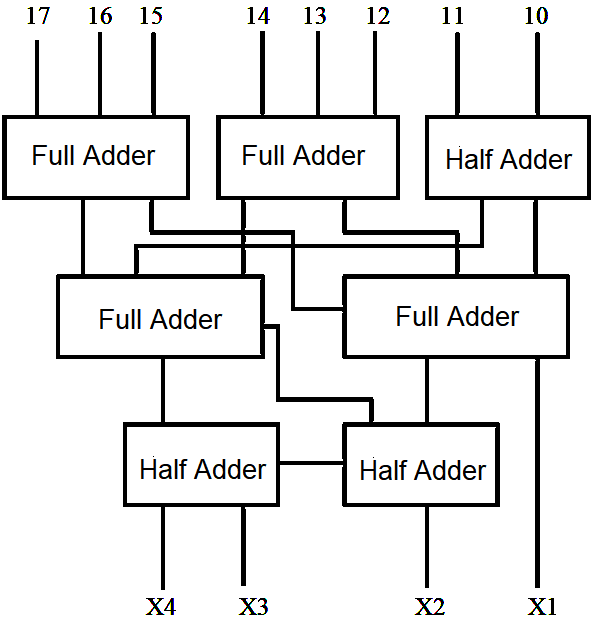}
\caption{8-to-4 Compressor Design}
\label{fig:8to4}
\end{figure}

Higher-order compressors, e.g., 5-to-3 (which compresses
five partial product bits into three) and 8-to-4 (which compresses eight partial product bits into four) \cite{M2}, allow us to construct high speed tree multipliers. Therefore, we also developed approximate FA based compressors, e.g., a 8-to-4 binary compressor is depicted in Figure \ref{fig:8to4}, for evaluation purposes. Table \ref{tab:Compressor} shows the power consumption and area for different approximate compressors implemented using approximate FAs. The area for approximate compressors exhibits a linear relationship with the area of FAs. However, it looks difficult to obtain a closed-form analytical expression for the power consumption. Few designs have a larger power consumption compared to the exact one, and this behavior needs more investigation. For that, and as a future work, we plan to use several approximate compressors with different approximation degrees in order to cover a larger design space.
Considering all options, the total combination of compressor settings grows exponentially O( (\# of FA designs)\textsuperscript{\# of FAs in compressor}) = O (11)\textsuperscript{4} = 14641 in our case. Therefore, to show the effectiveness of designing approximate compressors based on approximate FAs, we chose four FAs only. These FAs have superior designs metrics. The best approximate FA in terms of delay and PDP was \textit{AMA5}, and in terms of power and area was \textit{AXA2}. Also, the best FA with low error rate was \textit{InXA1}. \textit{AMA3} has moderate characteristics regarding area, power, delay, and number of errors. These selected FAs are used to design approximate high-order compressors, which in turn can be used for designing approximate tree multipliers. \textit{However, these selected compressors are not guaranteed to be the optimal ones. But, they exhibit some improvements compared to the exact designs.}

\begin{table}[!t]
  \centering
  \caption{Power Consumption and Area for Different Approximate Compressors based on Different Approximate FAs}
  \resizebox{0.99\textwidth}{!} {
    \begin{tabular}{|c|c|c|c|c|c|c|r|c|c|c|c|c|c|c|}
\cmidrule{2-7}\cmidrule{10-15}    \multicolumn{1}{r|}{} & \multicolumn{6}{c|}{Power Consumption ($\mu$w)for Different Compressors} & \multicolumn{1}{r}{} &       & \multicolumn{6}{c|}{Area (number of transistors)for Different Compressors} \\
\cmidrule{2-7}\cmidrule{10-15}    \multicolumn{1}{r|}{} & \multicolumn{6}{c|}{\cellcolor[rgb]{ .573,  .816,  .314} Compressor Type} & \multicolumn{1}{r}{} &       & \multicolumn{6}{c|}{\cellcolor[rgb]{ .573,  .816,  .314} Compressor Type} \\
\cmidrule{1-7}\cmidrule{9-15}    \rowcolor[rgb]{ 1,  1,  0} \textbf{FA Type} & \cellcolor[rgb]{ .573,  .816,  .314} \textbf{3-2} & \cellcolor[rgb]{ .573,  .816,  .314} \textbf{4-3} & \cellcolor[rgb]{ .573,  .816,  .314} \textbf{5-3} & \cellcolor[rgb]{ .573,  .816,  .314} \textbf{6-3} & \cellcolor[rgb]{ .573,  .816,  .314} \textbf{7-3} & \cellcolor[rgb]{ .573,  .816,  .314} \textbf{8-4} & \cellcolor[rgb]{ 1,  1,  1}  & \textbf{FA Type} & \cellcolor[rgb]{ .573,  .816,  .314} \textbf{3-2} & \cellcolor[rgb]{ .573,  .816,  .314} \textbf{4-3} & \cellcolor[rgb]{ .573,  .816,  .314} \textbf{5-3} & \cellcolor[rgb]{ .573,  .816,  .314} \textbf{6-3} & \cellcolor[rgb]{ .573,  .816,  .314} \textbf{7-3} & \cellcolor[rgb]{ .573,  .816,  .314} \textbf{8-4} \\
\cmidrule{1-7}\cmidrule{9-15}    \rowcolor[rgb]{ 1,  1,  0} \textbf{Exact} & \cellcolor[rgb]{ 1,  1,  1} \textbf{0.562} & \cellcolor[rgb]{ .851,  .851,  .851} \textbf{1.469} & \cellcolor[rgb]{ 1,  1,  1} \textbf{1.659} & \cellcolor[rgb]{ .851,  .851,  .851} \textbf{1.466} & \cellcolor[rgb]{ 1,  1,  1} \textbf{1.355} & \cellcolor[rgb]{ .851,  .851,  .851} \textbf{2.198} & \cellcolor[rgb]{ 1,  1,  1}  & \textbf{Exact} & \cellcolor[rgb]{ 1,  1,  1} \textbf{28} & \cellcolor[rgb]{ .851,  .851,  .851} \textbf{56} & \cellcolor[rgb]{ 1,  1,  1} \textbf{70} & \cellcolor[rgb]{ .851,  .851,  .851} \textbf{98} & \cellcolor[rgb]{ 1,  1,  1} \textbf{112} & \cellcolor[rgb]{ .851,  .851,  .851} \textbf{154} \\
\cmidrule{1-7}\cmidrule{9-15}    \rowcolor[rgb]{ 1,  1,  0} \textbf{M1} & \cellcolor[rgb]{ 1,  1,  1} \textbf{0.5474} & \cellcolor[rgb]{ .851,  .851,  .851} \textbf{0.9696} & \cellcolor[rgb]{ 1,  1,  1} \textbf{1.494} & \cellcolor[rgb]{ .851,  .851,  .851} \textbf{0.9258} & \cellcolor[rgb]{ 1,  1,  1} \textbf{1.138} & \cellcolor[rgb]{ .851,  .851,  .851} \textbf{1.65} & \cellcolor[rgb]{ 1,  1,  1}  & \textbf{M1} & \cellcolor[rgb]{ 1,  1,  1} \textbf{20} & \cellcolor[rgb]{ .851,  .851,  .851} \textbf{48} & \cellcolor[rgb]{ 1,  1,  1} \textbf{54} & \cellcolor[rgb]{ .851,  .851,  .851} \textbf{74} & \cellcolor[rgb]{ 1,  1,  1} \textbf{80} & \cellcolor[rgb]{ .851,  .851,  .851} \textbf{122} \\
\cmidrule{1-7}\cmidrule{9-15}    \rowcolor[rgb]{ 1,  1,  0} \textbf{M2} & \cellcolor[rgb]{ 1,  1,  1} \textbf{0.4525} & \cellcolor[rgb]{ .851,  .851,  .851} \textbf{1.224} & \cellcolor[rgb]{ 1,  1,  1} \textbf{1.189} & \cellcolor[rgb]{ 1,  .78,  .808} \textcolor[rgb]{ .612,  0,  .024}{\textbf{1.536}} & \cellcolor[rgb]{ 1,  1,  1} \textbf{1.321} & \cellcolor[rgb]{ .851,  .851,  .851} \textbf{1.609} & \cellcolor[rgb]{ 1,  1,  1}  & \textbf{M2} & \cellcolor[rgb]{ 1,  1,  1} \textbf{14} & \cellcolor[rgb]{ .851,  .851,  .851} \textbf{42} & \cellcolor[rgb]{ 1,  1,  1} \textbf{42} & \cellcolor[rgb]{ .851,  .851,  .851} \textbf{56} & \cellcolor[rgb]{ 1,  1,  1} \textbf{56} & \cellcolor[rgb]{ .851,  .851,  .851} \textbf{98} \\
\cmidrule{1-7}\cmidrule{9-15}    \rowcolor[rgb]{ 1,  1,  0} \textbf{M3} & \cellcolor[rgb]{ 1,  1,  1} \textbf{0.4489} & \cellcolor[rgb]{ .851,  .851,  .851} \textbf{0.6813} & \cellcolor[rgb]{ 1,  1,  1} \textbf{1.378} & \cellcolor[rgb]{ .851,  .851,  .851} \textbf{1.073} & \cellcolor[rgb]{ 1,  1,  1} \textbf{0.6157} & \cellcolor[rgb]{ .851,  .851,  .851} \textbf{0.9114} & \cellcolor[rgb]{ 1,  1,  1}  & \textbf{M3} & \cellcolor[rgb]{ 1,  1,  1} \textbf{11} & \cellcolor[rgb]{ .851,  .851,  .851} \textbf{39} & \cellcolor[rgb]{ 1,  1,  1} \textbf{36} & \cellcolor[rgb]{ .851,  .851,  .851} \textbf{47} & \cellcolor[rgb]{ 1,  1,  1} \textbf{44} & \cellcolor[rgb]{ .851,  .851,  .851} \textbf{86} \\
\cmidrule{1-7}\cmidrule{9-15}    \rowcolor[rgb]{ 1,  1,  0} \textbf{M4} & \cellcolor[rgb]{ 1,  1,  1} \textbf{0.5228} & \cellcolor[rgb]{ .851,  .851,  .851} \textbf{0.9988} & \cellcolor[rgb]{ 1,  1,  1} \textbf{1.176} & \cellcolor[rgb]{ .851,  .851,  .851} \textbf{1.183} & \cellcolor[rgb]{ 1,  1,  1} \textbf{1.037} & \cellcolor[rgb]{ .851,  .851,  .851} \textbf{1.449} & \cellcolor[rgb]{ 1,  1,  1}  & \textbf{M4} & \cellcolor[rgb]{ 1,  1,  1} \textbf{15} & \cellcolor[rgb]{ .851,  .851,  .851} \textbf{43} & \cellcolor[rgb]{ 1,  1,  1} \textbf{44} & \cellcolor[rgb]{ .851,  .851,  .851} \textbf{59} & \cellcolor[rgb]{ 1,  1,  1} \textbf{60} & \cellcolor[rgb]{ .851,  .851,  .851} \textbf{102} \\
\cmidrule{1-7}\cmidrule{9-15}    \rowcolor[rgb]{ 1,  1,  0} \textbf{M5} & \cellcolor[rgb]{ 1,  1,  1} \textbf{0.4802} & \cellcolor[rgb]{ .851,  .851,  .851} \textbf{0.9333} & \cellcolor[rgb]{ 1,  1,  1} \textbf{1.199} & \cellcolor[rgb]{ .851,  .851,  .851} \textbf{1.023} & \cellcolor[rgb]{ 1,  1,  1} \textbf{0.8753} & \cellcolor[rgb]{ .851,  .851,  .851} \textbf{1.791} & \cellcolor[rgb]{ 1,  1,  1}  & \textbf{M5} & \cellcolor[rgb]{ 1,  1,  1} \textbf{8} & \cellcolor[rgb]{ .851,  .851,  .851} \textbf{36} & \cellcolor[rgb]{ 1,  1,  1} \textbf{30} & \cellcolor[rgb]{ .851,  .851,  .851} \textbf{38} & \cellcolor[rgb]{ 1,  1,  1} \textbf{32} & \cellcolor[rgb]{ .851,  .851,  .851} \textbf{74} \\
\cmidrule{1-7}\cmidrule{9-15}    \rowcolor[rgb]{ 1,  1,  0} \textbf{X1} & \cellcolor[rgb]{ 1,  1,  1} \textbf{0.4511} & \cellcolor[rgb]{ 1,  .78,  .808} \textcolor[rgb]{ .612,  0,  .024}{\textbf{1.586}} & \cellcolor[rgb]{ 1,  1,  1} \textbf{1.128} & \cellcolor[rgb]{ 1,  .78,  .808} \textcolor[rgb]{ .612,  0,  .024}{\textbf{1.562}} & \cellcolor[rgb]{ 1,  .78,  .808} \textcolor[rgb]{ .612,  0,  .024}{\textbf{1.521}} & \cellcolor[rgb]{ .851,  .851,  .851} \textbf{2.142} & \cellcolor[rgb]{ 1,  1,  1}  & \textbf{X1} & \cellcolor[rgb]{ 1,  1,  1} \textbf{8} & \cellcolor[rgb]{ .851,  .851,  .851} \textbf{36} & \cellcolor[rgb]{ 1,  1,  1} \textbf{30} & \cellcolor[rgb]{ .851,  .851,  .851} \textbf{38} & \cellcolor[rgb]{ 1,  1,  1} \textbf{32} & \cellcolor[rgb]{ .851,  .851,  .851} \textbf{74} \\
\cmidrule{1-7}\cmidrule{9-15}    \rowcolor[rgb]{ 1,  1,  0} \textbf{X2} & \cellcolor[rgb]{ 1,  1,  1} \textbf{0.4584} & \cellcolor[rgb]{ .851,  .851,  .851} \textbf{1.296} & \cellcolor[rgb]{ 1,  1,  1} \textbf{1.563} & \cellcolor[rgb]{ .851,  .851,  .851} \textbf{0.8141} & \cellcolor[rgb]{ 1,  1,  1} \textbf{0.7489} & \cellcolor[rgb]{ 1,  .78,  .808} \textcolor[rgb]{ .612,  0,  .024}{\textbf{2.77}} & \cellcolor[rgb]{ 1,  1,  1}  & \textbf{X2} & \cellcolor[rgb]{ 1,  1,  1} \textbf{6} & \cellcolor[rgb]{ .851,  .851,  .851} \textbf{34} & \cellcolor[rgb]{ 1,  1,  1} \textbf{26} & \cellcolor[rgb]{ .851,  .851,  .851} \textbf{32} & \cellcolor[rgb]{ 1,  1,  1} \textbf{24} & \cellcolor[rgb]{ .851,  .851,  .851} \textbf{66} \\
\cmidrule{1-7}\cmidrule{9-15}    \rowcolor[rgb]{ 1,  1,  0} \textbf{X3} & \cellcolor[rgb]{ 1,  1,  1} \textbf{0.3544} & \cellcolor[rgb]{ .851,  .851,  .851} \textbf{1.349} & \cellcolor[rgb]{ 1,  1,  1} \textbf{1.429} & \cellcolor[rgb]{ .851,  .851,  .851} \textbf{1.231} & \cellcolor[rgb]{ 1,  1,  1} \textbf{0.4742} & \cellcolor[rgb]{ 1,  .78,  .808} \textcolor[rgb]{ .612,  0,  .024}{\textbf{2.316}} & \cellcolor[rgb]{ 1,  1,  1}  & \textbf{X3} & \cellcolor[rgb]{ 1,  1,  1} \textbf{8} & \cellcolor[rgb]{ .851,  .851,  .851} \textbf{36} & \cellcolor[rgb]{ 1,  1,  1} \textbf{30} & \cellcolor[rgb]{ .851,  .851,  .851} \textbf{38} & \cellcolor[rgb]{ 1,  1,  1} \textbf{32} & \cellcolor[rgb]{ .851,  .851,  .851} \textbf{74} \\
\cmidrule{1-7}\cmidrule{9-15}    \rowcolor[rgb]{ 1,  1,  0} \textbf{In1} & \cellcolor[rgb]{ 1,  1,  1} \textbf{0.1823} & \cellcolor[rgb]{ 1,  .78,  .808} \textcolor[rgb]{ .612,  0,  .024}{\textbf{1.569}} & \cellcolor[rgb]{ 1,  1,  1} \textbf{0.9423} & \cellcolor[rgb]{ .851,  .851,  .851} \textbf{0.4842} & \cellcolor[rgb]{ 1,  1,  1} \textbf{1.296} & \cellcolor[rgb]{ 1,  .78,  .808} \textcolor[rgb]{ .612,  0,  .024}{\textbf{2.413}} & \cellcolor[rgb]{ 1,  1,  1}  & \textbf{In1} & \cellcolor[rgb]{ 1,  1,  1} \textbf{6} & \cellcolor[rgb]{ .851,  .851,  .851} \textbf{34} & \cellcolor[rgb]{ 1,  1,  1} \textbf{26} & \cellcolor[rgb]{ .851,  .851,  .851} \textbf{32} & \cellcolor[rgb]{ 1,  1,  1} \textbf{24} & \cellcolor[rgb]{ .851,  .851,  .851} \textbf{66} \\
\cmidrule{1-7}\cmidrule{9-15}    \rowcolor[rgb]{ 1,  1,  0} \textbf{In2} & \cellcolor[rgb]{ 1,  1,  1} \textbf{0.5018} & \cellcolor[rgb]{ .851,  .851,  .851} \textbf{1.324} & \cellcolor[rgb]{ 1,  .78,  .808} \textcolor[rgb]{ .612,  0,  .024}{\textbf{2.114}} & \cellcolor[rgb]{ .851,  .851,  .851} \textbf{0.3441} & \cellcolor[rgb]{ 1,  1,  1} \textbf{0.8087} & \cellcolor[rgb]{ 1,  .78,  .808} \textcolor[rgb]{ .612,  0,  .024}{\textbf{3.844}} & \cellcolor[rgb]{ 1,  1,  1}  & \textbf{In2} & \cellcolor[rgb]{ 1,  1,  1} \textbf{8} & \cellcolor[rgb]{ .851,  .851,  .851} \textbf{36} & \cellcolor[rgb]{ 1,  1,  1} \textbf{30} & \cellcolor[rgb]{ .851,  .851,  .851} \textbf{38} & \cellcolor[rgb]{ 1,  1,  1} \textbf{32} & \cellcolor[rgb]{ .851,  .851,  .851} \textbf{74} \\
\cmidrule{1-7}\cmidrule{9-15}    \rowcolor[rgb]{ 1,  1,  0} \textbf{In3} & \cellcolor[rgb]{ 1,  1,  1} \textbf{0.5504} & \cellcolor[rgb]{ .851,  .851,  .851} \textbf{1.345} & \cellcolor[rgb]{ 1,  1,  1} \textbf{1.234} & \cellcolor[rgb]{ .851,  .851,  .851} \textbf{0.7866} & \cellcolor[rgb]{ 1,  .78,  .808} \textcolor[rgb]{ .612,  0,  .024}{\textbf{1.636}} & \cellcolor[rgb]{ 1,  .78,  .808} \textcolor[rgb]{ .612,  0,  .024}{\textbf{5.27}} & \cellcolor[rgb]{ 1,  1,  1}  & \textbf{In3} & \cellcolor[rgb]{ 1,  1,  1} \textbf{6} & \cellcolor[rgb]{ .851,  .851,  .851} \textbf{34} & \cellcolor[rgb]{ 1,  1,  1} \textbf{26} & \cellcolor[rgb]{ .851,  .851,  .851} \textbf{32} & \cellcolor[rgb]{ 1,  1,  1} \textbf{24} & \cellcolor[rgb]{ .851,  .851,  .851} \textbf{66} \\
\cmidrule{1-7}\cmidrule{9-15}    
\end{tabular}%
}
  \label{tab:Compressor}%
\end{table}%


\section{\textbf{Multiplier Basic Blocks}} \label{sec:8X8}

In this section, we use the approximate FAs and compressors, described above, to design 8x8 array and tree based multipliers, respectively. These 8x8 approximate multipliers will act as our basic blocks for designing higher-order multipliers, i.e., 32x32 and 64x64, as it will be discussed in Section \ref{sec:MainModule}.

\subsection{\textbf{8x8 Array Multiplier}}  \label{sec:Array8}
 An n-bit array multiplier \cite{rabaey} is composed of \textit{n\textsuperscript{2}} AND gates for partial products generation, and \textit{n-1} n-bit adders for partial products accumulation. The design space of an \textit{n}x\textit{n} approximate array multiplier is quite huge, since it depends on the type of FA used in the array, and the \textit{number} of approximate FAs (from 0 to \textit{n}) used in the n-bit adder. Considering all options, the total combination of multiplier settings grow exponentially O( (\# of FAs)\textsuperscript{MultiplierSize\textsuperscript{2}}) = O ((11)\textsuperscript{n\textsuperscript{2}}) = (11)\textsuperscript{64} in our case.

We have used all 11 Pareto approximate FAs, described in Section \ref{sec:ApproxFA}, to construct 8x8 approximate array multipliers, based on only one FA type per design to avoid the exponential growth of the design space. Regarding the degree of approximation, we have used two options: i) all FAs are approximate, and ii) FAs that contribute to the least significant 50\% of the resultant bits are approximated to maintain acceptable accuracy as recommended by \cite{Vaibhav} \cite{Reddy} \cite{Shao}. Thus, we have designed, evaluated and compared 22 different options for building 8x8 approximate array multipliers as shown in Table \ref{table:8X8Array}, using the TSMC65nm technology. The type of the multiplier in Table \ref{table:8X8Array} consists of two parts, i.e., the name of the adder used for the most significant and least significant part. For example, in EM1, the most significant part is based on an exact (E) adder and the least significant part is based on the mirror adder 1 (M1).

\begin{table}[!t]
  \centering
  \caption{8x8 Approximate Array Multiplier}
  \scalebox{0.7}{  
    \begin{tabular}{|c|c|c|c|c|c|c|c|}
    \toprule
          \textbf{Type} & \multicolumn{1}{p{4.055em}|}{\textbf{MRED}} & \multicolumn{1}{p{4.055em}|}{\textbf{MED}} & \multicolumn{1}{p{4.055em}|}{\textbf{ER}} & \multicolumn{1}{p{4.055em}|}{\textbf{NMED}} & \multicolumn{1}{p{2.555em}|}{\textbf{Delay (ps)}} & \multicolumn{1}{p{2.78em}|}{\textbf{Power ($\mu$W)}} & \multicolumn{1}{p{2.11em}|}{\textbf{size}} \\
    \midrule
    \textbf{EE} & 00  & 00   & 00  & 00   & 527   & 31.41 & 1456 \\
    \midrule
    \textbf{EM1} & 8.55E-02 & 2.55E+02 & 9.70E-01 & 3.93E-03 & 527   & 24.17 & 1288 \\
    \midrule
    \textbf{M1M1} & 2.13E+00 & 1.33E+04 & 9.96E-01 & 2.05E-01 & 865   & 14.75 & 1072 \\
    \midrule
    \textbf{EM2} & 1.85E-01 & 2.29E+02 & 9.90E-01 & 3.52E-03 & 557   & 22.97 & 1162 \\
    \midrule
    \textbf{M2M2} & 1.73E+01 & 1.68E+04 & 1.00E+00 & 2.58E-01 & 600   & 14.4  & 784 \\
    \midrule
    \textbf{EM3} & 4.03E-01 & 4.72E+02 & 9.99E-01 & 7.26E-03 & 605   & 24.95 & 1099 \\
    \midrule
    \textbf{M3M3} & 1.25E+01 & 1.72E+04 & 1.00E+00 & 2.64E-01 & 598   & 15.31 & 640 \\
    \midrule
    \textbf{EM4} & 3.64E-02 & 1.11E+02 & 9.70E-01 & 1.71E-03 & 573   & 21.85 & 1183 \\
    \midrule
    \textbf{M4M4} & 6.11E-01 & 6.41E+03 & 9.96E-01 & 9.86E-02 & 313   & 11.17 & 832 \\
    \midrule
    \textbf{EM5} & 3.03E-02 & 1.01E+02 & 9.30E-01 & 1.56E-03 & 573   & 22.15 & 1036 \\
    \midrule
    \textbf{M5M5} & 6.76E-01 & 8.24E+03 & 9.90E-01 & 1.27E-01 & 250   & 10.69 & 496 \\
    \midrule
    \textbf{EX1} & 1.18E-01 & 2.09E+02 & 9.71E-01 & 3.21E-03 & 546   & 31.86 & 1036 \\
    \midrule
    \textbf{X1X1} & 2.84E+00 & 1.05E+04 & 9.96E-01 & 1.61E-01 & 558   & 21.33 & 496 \\
    \midrule
    \textbf{EX2} & 1.09E-01 & 1.88E+02 & 1.00E+00 & 2.89E-03 & 569   & 23.38 & 994 \\
    \midrule
    \textbf{X2X2} & 1.18E+01 & 1.51E+04 & 1.00E+00 & 2.31E-01 & 250   & 13.91 & 400 \\
    \midrule
    \textbf{EX3} & 7.96E-02 & 3.48E+02 & 6.15E-01 & 5.35E-03 & 536   & 25.54 & 1036 \\
    \midrule
    \textbf{X3X3} & 9.88E-01 & 1.63E+04 & 9.96E-01 & 2.50E-01 & 197   & 15.06 & 496 \\
    \midrule
    \textbf{EIn1} & 7.50E-02 & 3.19E+02 & 6.15E-01 & 4.91E-03 & 517   & 26.07 & 994 \\
    \midrule
    \textbf{In1In1} & 1.62E+00 & 1.02E+04 & 8.54E-01 & 1.56E-01 & 403   & 14.82 & 400 \\
    \midrule
    \textbf{EIn2} & 3.68E-02 & 1.80E+02 & 5.84E-01 & 2.76E-03 & 528   & 28.79 & 1036 \\
    \midrule
    \textbf{In2In2} & 4.63E-01 & 8.28E+03 & 8.26E-01 & 1.27E-01 & 340   & 12.56 & 496 \\
    \midrule
    \textbf{EIn3} & 1.85E-01 & 2.29E+02 & 9.90E-01 & 3.52E-03 & 556   & 27.96 & 994 \\
    \midrule
    \textbf{In3In3} & 1.73E+01 & 1.68E+04 & 1.00E+00 & 2.58E-01 & 404   & 23.92 & 400 \\
    \bottomrule
    \end{tabular}%
    }
  \label{table:8X8Array}
\end{table}%

\begin{figure}[!h]
 \centering
\includegraphics[width=1.0\textwidth , height=5cm]{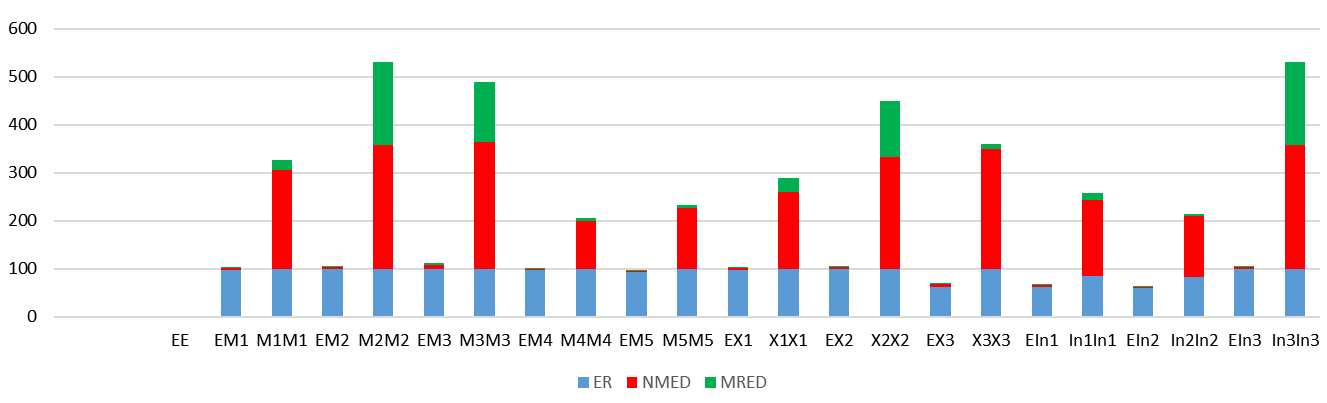}
\caption{ER x$10^{-2}$, NMED x$10^{-3}$ and MRED x$10^{-1}$ of 8x8 Array Multiplier}
\label{fig:error_8x8Array}
\end{figure}

 For our approximate designs, a specific approximation degree, from 1 to \textit{2n}, rather than \textit{n}, may be chosen based on the maximum error allowed for a specific application, where in\cite{SALSA} \cite{SASIMI}, it is mentioned that it is suitable to chose a value of 10\% for Maximum ED and 0.5\% for MED. Figure \ref{fig:error_8x8Array} shows the ER, NMED and MRED for various 8x8 array multipliers. It is clear that fully approximate multipliers have high NMED. \textit{EM5} has the lowest NMED, and \textit{EM4} has a close-to-minimal NMED. Designs with high NMED have a high MRED too. It can be observed that \textit{EIn2} exhibits the lowest ER. Also, \textit{EX3} and \textit{EIn1} have the same close-to-minimal ER. 

 \begin{figure}[!h]
 \centering
\includegraphics[width=1.0\textwidth, height=5cm]{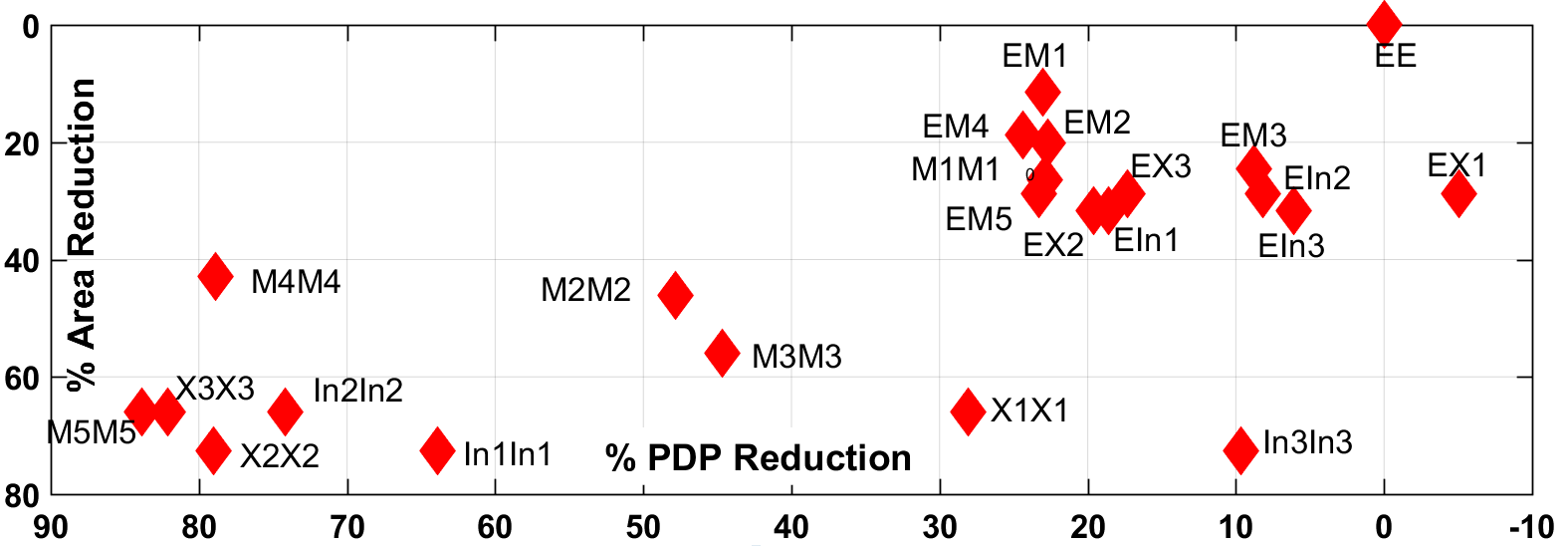}
\caption{Area and PDP Reduction of 8x8 Array Multiplier}
\label{fig:PDP_Area_NMED_8x8Array}
\end{figure}

 As shown in Table \ref{table:8X8Array} and Figure \ref{fig:error_8x8Array}, there is no single design that is superior in all design metrics. Therefore, a Pareto-analysis for the improvements in area and PDP is shown for the different  designs proposed throughout this work. \textit{X3X3} has the lowest delay, and \textit{M5M5} and \textit{X2X2} also exhibit a low delay. \textit{M5M5}, \textit{M4M4} and \textit{In2In2} exhibit the lowest power consumption among the 22 different designs. The size of the approximate multiplier exhibits a linear relationship with the degree of approximation. Thus, \textit{X2X2}, \textit{In1In1} and \textit{In3In3}, have the smallest size.
   

Figure \ref{fig:PDP_Area_NMED_8x8Array} shows the area and PDP reduction of 8x8 array multipliers. The best designs are located on the bottom left corner. \textit{M5M5} is a Pareto-design with PDP reduction of 84\% and area reduction of 65\%. The design \textit{X3X3} is Non-Pareto because it has the same area reduction as the \textit{M5M5} but with a smaller PDP reduction. However, we have to consider other \textit{error metrics}. Some designs such as \textit{EX1} have increased PDP due to excessive switching activity compared to the original design.

\subsection{\textbf{8x8 Tree Multiplier}} \label{sec:Tree8}

The Wallace multiplier\cite{Parhami} is an efficient parallel multiplier that is composed of a tree of half adders (HAs) and FAs. The main idea is that, the adders in each layer operate in parallel without carry propagation until the generation of two rows of partial products. The design space for approximate 8x8 tree multipliers \cite{Parhami} is also quite large, depending on the \textit{compressor type} and \textit{approximation degree}. To avoid the exponentially growing design space, we choose to use compressors of the same type in the multiplier design. Also, we use two options for approximation degree: i) all compressors are approximate, and ii) compressors that contribute to the lowest significant 50\% of the resultant bits are approximated to maintain an acceptable accuracy. Thus, based on the four shortlisted compressors, explained in Section \ref{sec:ApproxFA}, we compared 8 options for approximate 8x8 tree multipliers and the results are given in Table \ref{table:8X8Tree}. The name of the multiplier consists of three parts. For example, CEM1 represents a compressor based multiplier (C), where the most significant part is based on an exact (E) compressor and the least significant part is composed of the mirror adder 1 (M1) based compressor. As shown in Table \ref{table:8X8Tree}, there is no single design superior is all metrics, but some designs are the best wrt some few metrics. 

\begin{figure}[h!]
\centering
\includegraphics[width=1.0\textwidth, height=5cm]{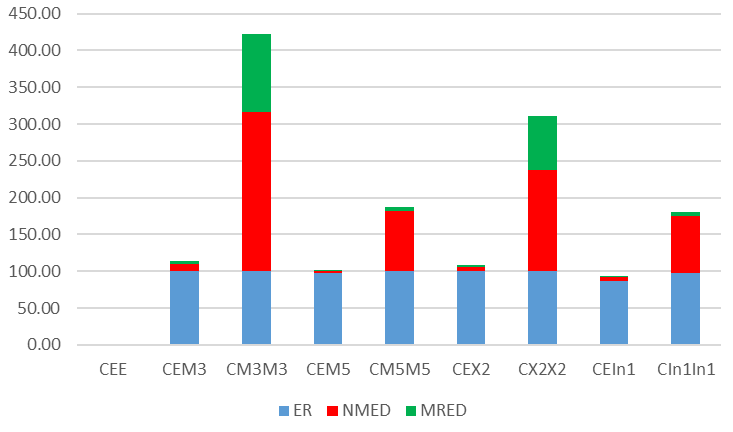}
\caption{ER x$10^{-2}$, NMED x$10^{-3}$ and MRED x$10^{-1}$ of 8x8 Tree Multiplier}
\label{fig:error_8x8Tree}
\end{figure}

Figure \ref{fig:error_8x8Tree} shows the ER, NMED and MRED for various 8x8 tree multipliers. Fully approximate designs have higher NMED and MRED than partially approximate designs. The designs based on \textit{InX1} (\textit{CEIn1} and \textit{CIn1In1}) exhibit the lowest ER. \textit{CM3M3} have the highest MRED. The designs based on \textit{AMA5} have the lowest delay and power consumption due to their simple structures. 

\begin{figure}[h!]
\centering
\includegraphics[width=1.0\textwidth, height=5cm]{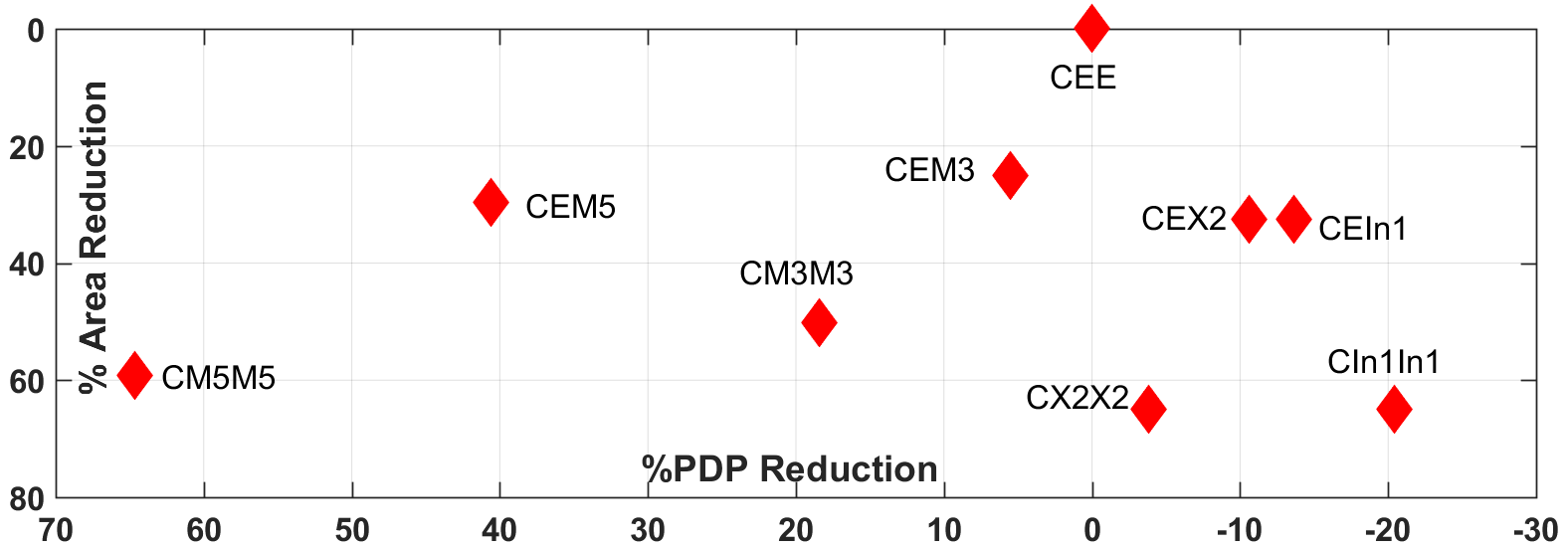}
\caption{Area and PDP Reduction of 8x8 Tree Multiplier}
\label{fig:PDP_Area_NMED_8x8Tree}
\end{figure}

As depicted in Figure \ref{fig:PDP_Area_NMED_8x8Tree} which shows area and PDP reduction, the best designs are on the left bottom corner, i.e., \textit{CM5M5} is a Pareto-design with maximum area and maximum PDP reduction. However, \textit{CEM5} is a non Pareto-design because it has less reduction. Few designs on the right side of the figure have a PDP value greater than the exact design, which makes them unsuitable for low-power design usage.

\begin{table}[!t]
\centering
 \caption{8x8 Approximate Tree Multiplier}
  \scalebox{0.7}{
    \begin{tabular}{|c|c|c|c|c|c|c|c|}
    \toprule
             \textbf{Type} & \multicolumn{1}{p{4.055em}|}{\textbf{MRED}} & \multicolumn{1}{p{4.055em}|}{\textbf{MED}} & \multicolumn{1}{p{4.055em}|}{\textbf{ER}} & \multicolumn{1}{p{4.055em}|}{\textbf{NMED}} & \multicolumn{1}{p{2.555em}|}{\textbf{Delay (ps)}} & \multicolumn{1}{p{2.78em}|}{\textbf{Power ($\mu$W)}} & \multicolumn{1}{p{2.11em}|}{\textbf{size}} \\
             
    \midrule
    \textbf{CEE} & 00 & 00 & 00 & 00 & 508   & 21.98 & 1218 \\
    \midrule
    \textbf{CEM3} & 4.76E-01 & 6.05E+02 & 1.00E+00 & 9.30E-03 & 537   & 19.65 & 912 \\
    \midrule
    \textbf{CM3M3} & 1.06E+01 & 1.41E+04 & 1.00E+00 & 2.16E-01 & 560   & 16.27 & 606 \\
    \midrule
    \textbf{CEM5} & 4.76E-02 & 1.54E+02 & 9.79E-01 & 2.40E-03 & 356   & 18.63 & 858 \\
    \midrule
    \textbf{CM5M5} & 5.16E-01 & 5.32E+03 & 9.99E-01 & 8.18E-02 & 282   & 13.99 & 498 \\
    \midrule
    \textbf{CEX2} & 3.28E-01 & 3.68E+02 & 9.97E-01 & 5.70E-03 & 525   & 23.52 & 822 \\
    \midrule
    \textbf{CX2X2} & 7.35E+00 & 8.95E+03 & 1.00E+00 & 1.38E-01 & 513   & 22.6  & 426 \\
    \midrule
    \textbf{CEIn1} & 9.03E-02 & 3.10E+02 & 8.73E-01 & 4.80E-03 & 505   & 25.12 & 822 \\
    \midrule
    \textbf{CIn1In1} & 5.08E-01 & 5.08E+03 & 9.75E-01 & 7.81E-02 & 500   & 26.89 & 426 \\
    \bottomrule
\hline
\end{tabular}  }
\label{table:8X8Tree}
\end{table}

\section{\textbf{ Higher-Order Multiplier Configuration}}   \label{sec:MainModule}

\begin{figure}[h!]
\centering
\includegraphics[width=6cm, height=3cm]{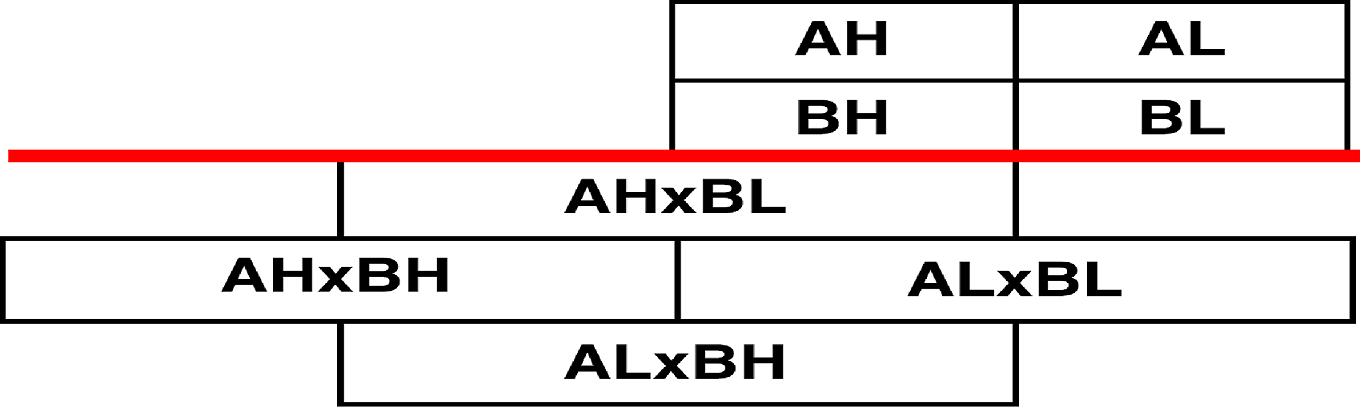}
\caption{16x16 Multiplier}
\label{fig:recursive}
\end{figure}

The 8x8 multiplier basic modules can be used to construct higher-order target multiplier modules. In this report, we use the example of designing a 16x16 multiplier to illustrate this process. The partial product tree of the 16x16 multiplication can be broken down into four products of 8x8 modules, which can be executed concurrently, as shown in Figure \ref{fig:recursive}.

In the case of high requirements of accuracy, an exact 8x8  multiplier can be used for the three most significant products, i.e., AHxBH, AHxBL, and ALxBH, and any one of the approximate designs can be used for the least significant product, i.e., ALxBL. For low accuracy requirements, only one 8x8 exact multiplier can be used for the most significant product, i.e., AHxBH, and any of the other approximate designs can be used for the three least significant products, i.e., AHxBL, ALxBH, and ALxBL. Modules that contribute to the lowest significant 50\% of the resultant bits are approximated to maintain accuracy as recommended by \cite{Vaibhav} \cite{Reddy} \cite{Shao}\cite{FEMTO}.

We choose to design 16x16 multipliers with an exact AHxBH multiplier, and with exact MSBs and approximate LSBs for AHxBL and ALxBH, and a fully approximate or approximate in LSBs only ALxBL. Any other approximation degree can be found based on the required quality function (maximum error, area, power or delay).
Therefore, when the 16x16 multipliers are explained, the types of AHxBH, AHxBL and ALxBH are eliminated  from the name, and only the type of ALxBL is used in the name of the multiplier.


\subsection{\textbf{16x16 Array Multiplier}} \label{sec:Array16}

\begin{figure}[h!]
\centering
\includegraphics[width=1.0\textwidth, height=5cm]{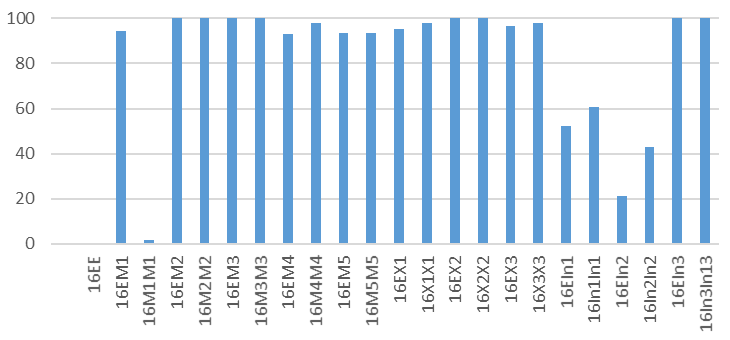}
\caption{ER x$10^{-2}$ of 16x16 Array Multiplier}
\label{fig:ER_16x16Array}
\end{figure}

\begin{figure}[h!]
\centering
\includegraphics[width=1.0\textwidth, height=5cm]{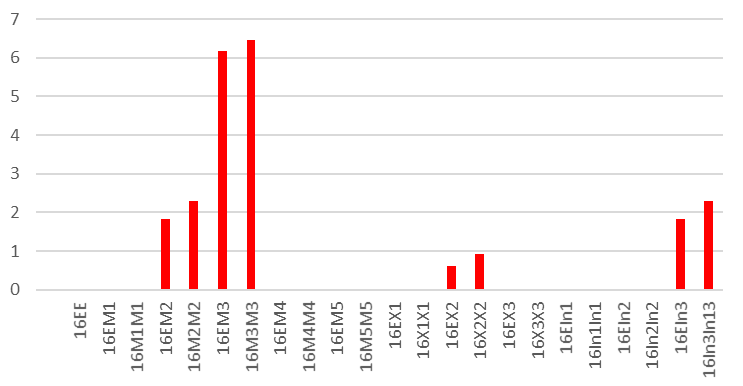}
\caption{NMED x$10^{-5}$ of 16x16 Array Multiplier}
\label{fig:NMED_16x16Array}
\end{figure}

\begin{figure}[h!]
\centering
\includegraphics[width=1.0\textwidth, height=5cm]{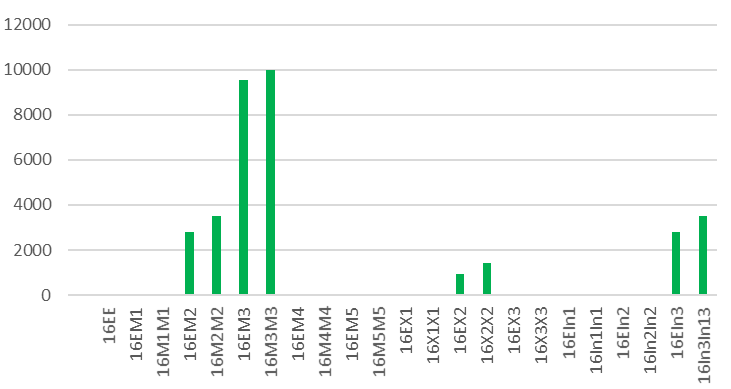}
\caption{MRED x$10^{-1}$of 16x16 Array Multiplier}
\label{fig:MRED_16x16Array}
\end{figure}

Table \ref{table:16X16Array} shows the simulation results for 16x16 approximate array multipliers, which shows similarities with Table \ref{table:8X8Array}. The multiplier name is based on the type of ALxBL module. Figures \ref{fig:ER_16x16Array}, \ref{fig:NMED_16x16Array} and \ref{fig:MRED_16x16Array} show the ER, NMED and MRED for 16x16 array multipliers, respectively. It can be observed that \textit{16M1M1} is the most accurate design with the lowest ER and lowest NMED. \textit{16EIn2} is the second accurate design with low ER and NMED. For NMED, the best designs are \textit{16M1M1}, \textit{16EIn2} and \textit{16In2In2}. Designs with high NMED show a high MRED value. \textit{EIn1In1} and \textit{16In3In3} have the lowest delay. Fully approximate designs exhibit the minimal delay. Generally, designs based on approximate mirror adders have the lowest power consumption, due to the elimination of static power dissipation. Since, the design size grows linearly with the FA size, fully approximate designs based on 6 transistors  cells including \textit{16X2X2}, \textit{16In1In1} and \textit{16In3In3} have the smallest number of transistors. Also, fully approximate designs including \textit{16M5M5}, \textit{16X1X1}, \textit{16X3X3} and \textit{16In2In2}, based on 8 transistors FAs, have a very small size as well.  Finally, the best designs regarding area reduction are \textit{16In1In1}, \textit{16X2X2} and \textit{16In3In3}.

\begin{table}[t!]
 \centering
 \caption{16x16 Approximate Array Multiplier}
   \scalebox{0.7}{ 
    \begin{tabular}{|c|c|c|c|c|c|c|c|}
    \toprule
       \textbf{Type} & \multicolumn{1}{p{4.055em}|}{\textbf{MRED}} & \multicolumn{1}{p{4.055em}|}{\textbf{MED}} & \multicolumn{1}{p{4.055em}|}{\textbf{ER}} & \multicolumn{1}{p{4.055em}|}{\textbf{NMED}} & \multicolumn{1}{p{2.555em}|}{\textbf{Delay (ps)}} & \multicolumn{1}{p{2.78em}|}{\textbf{Power ($\mu$W)}} & \multicolumn{1}{p{2.11em}|}{\textbf{size}} \\
    \midrule
    \textbf{16EE} & 00 & 00 & 00 & 00 & 514   & 156.8 & 5824 \\
    \midrule
    \textbf{16EM1} & 1.19E-02 & 6.31E+04 & 9.44E-01 & 7.69E-10 & 534   & 130.1 & 5320 \\
    \midrule
    \textbf{16M1M1} & 1.71E-04 & 1.33E+03 & 1.76E-02 & 1.10E-11 & 526   & 118.4 & 5104 \\
    \midrule
    \textbf{16EM2} & 2.82E+02 & 1.14E+05 & 1.00E+00 & 1.82E-05 & 533   & 128.4 & 4942 \\
    \midrule
    \textbf{16M2M2} & 3.53E+02 & 1.33E+05 & 1.00E+00 & 2.28E-05 & 477   & 116.5 & 4562 \\
    \midrule
    \textbf{16EM3} & 9.53E+02 & 3.34E+05 & 1.00E+00 & 6.16E-05 & 519   & 131.6 & 4753 \\
    \midrule
    \textbf{16M3M3} & 9.98E+02 & 3.51E+05 & 1.00E+00 & 6.45E-05 & 490   & 120.4 & 4294 \\
    \midrule
    \textbf{16EM4} & 7.80E-03 & 3.36E+04 & 9.29E-01 & 5.04E-10 & 522   & 118.8 & 5005 \\
    \midrule
    \textbf{16M4M4} & 7.90E-03 & 3.32E+04 & 9.79E-01 & 5.11E-10 & 506   & 105.1 & 4654 \\
    \midrule
    \textbf{16EM5} & 8.20E-03 & 4.06E+04 & 9.34E-01 & 5.30E-10 & 533   & 119   & 4564 \\
    \midrule
    \textbf{16M5M5} & 8.20E-03 & 4.06E+04 & 9.34E-01 & 5.30E-10 & 535   & 105.1 & 4024 \\
    \midrule
    \textbf{16EX1} & 1.15E-02 & 5.22E+04 & 9.51E-01 & 7.43E-10 & 513   & 154.9 & 4564 \\
    \midrule
    \textbf{16X1X1} & 1.29E-02 & 5.74E+04 & 9.79E-01 & 8.34E-10 & 520   & 138.5 & 4024 \\
    \midrule
    \textbf{16EX2} & 9.40E+01 & 5.96E+04 & 1.00E+00 & 6.07E-06 & 521   & 138   & 4438 \\
    \midrule
    \textbf{16X2X2} & 1.41E+02 & 6.85E+04 & 1.00E+00 & 9.11E-06 & 514   & 127.4 & 3844 \\
    \midrule
    \textbf{16EX3} & 1.69E-02 & 9.09E+04 & 9.65E-01 & 1.09E-09 & 515   & 134   & 4564 \\
    \midrule
    \textbf{16X3X3} & 1.97E-02 & 1.05E+05 & 9.79E-01 & 1.27E-09 & 518   & 121.8 & 4024 \\
    \midrule
    \textbf{16EIn1} & 7.80E-03 & 4.57E+04 & 5.24E-01 & 5.04E-10 & 519   & 134.2 & 4438 \\
    \midrule
    \textbf{16In1In1} & 8.40E-03 & 4.93E+04 & 6.09E-01 & 5.43E-10 & 408   & 121.7 & 3844 \\
    \midrule
    \textbf{16EIn2} & 1.60E-03 & 8.08E+03 & 2.14E-01 & 1.03E-10 & 537   & 146.9 & 4564 \\
    \midrule
    \textbf{16In2In2} & 2.20E-03 & 1.13E+04 & 4.29E-01 & 1.42E-10 & 500   & 126.4 & 4024 \\
    \midrule
    \textbf{16EIn3} & 2.82E+02 & 1.14E+05 & 1.00E+00 & 1.82E-05 & 527   & 157.6 & 4438 \\
    \midrule
    \textbf{16In3In13} & 3.53E+02 & 1.33E+05 & 1.00E+00 & 2.28E-05 & 412   & 153.2 & 3844 \\
    \bottomrule

 \end{tabular} 
  }
\label{table:16X16Array}
\end{table}



\begin{figure}[h!]
\centering
\includegraphics[width=1.0\textwidth, height=5cm]{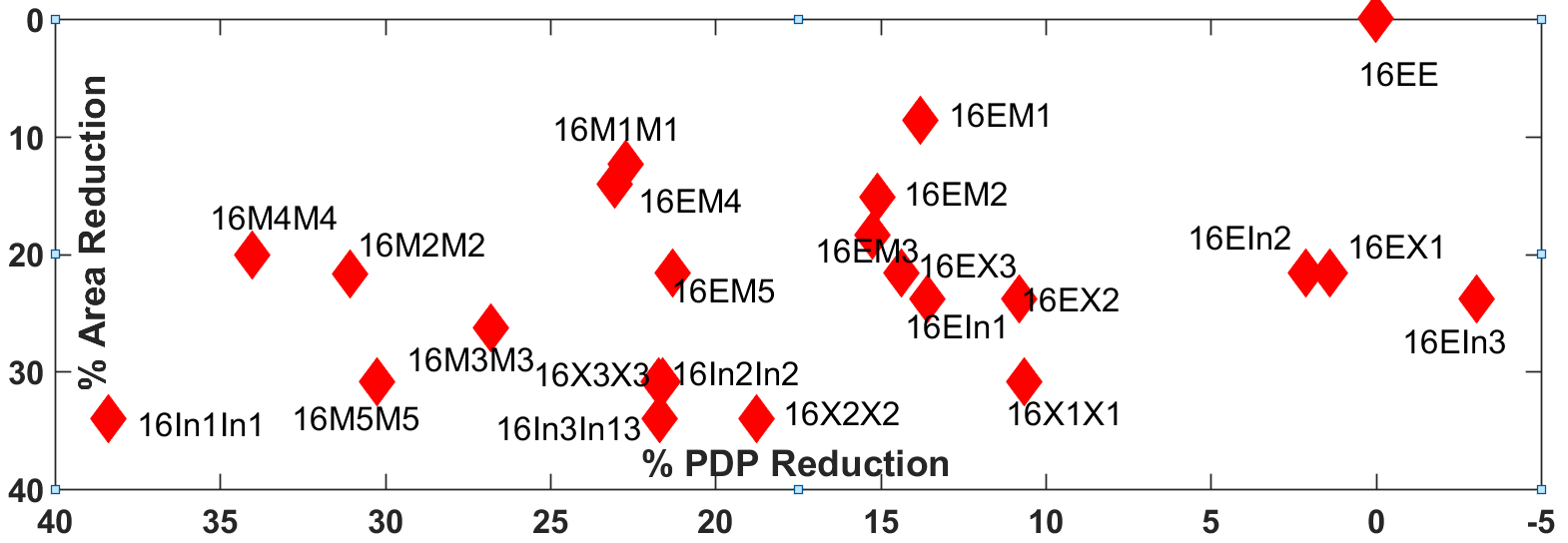}
\caption{Area and PDP Reduction of 16x16 Array Multiplier}
\label{fig:PDP_Area_NMED_16x16Array}
\end{figure}

As depicted in Figure \ref{fig:PDP_Area_NMED_16x16Array} which shows the reduction in area and PDP for 16x16 array multipliers, the best designs are on the lower left corner, i.e., \textit{16In1In1} and \textit{16In3In3} are Pareto-designs while \textit{16M4M4} is a non-Pareto design. Designs with negative PDP reduction values, indicate a power or delay larger than the exact design.

\subsection{\textbf{16x16 Tree Multiplier}}  \label{sec:Tree16}

\begin{figure}[h!]
\centering
\includegraphics[width=1.0\textwidth, height=5cm]{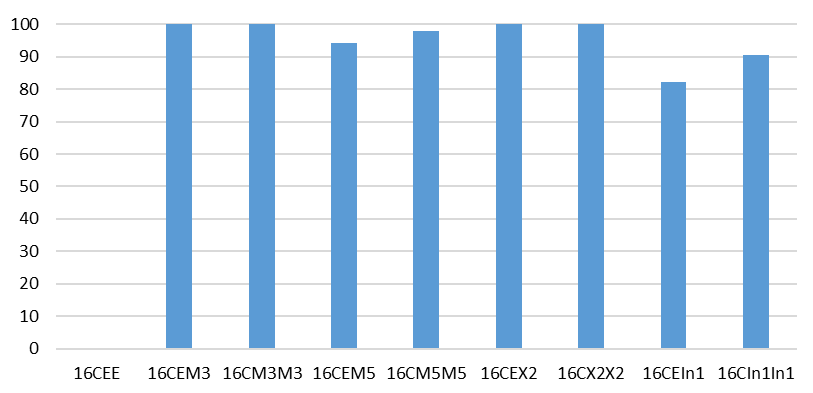}
\caption{ER x$10^{-2}$ of 16x16 Tree Multiplier}
\label{fig:ER_16x16Tree}
\end{figure}

\begin{figure}[h!]
\centering
\includegraphics[width=1.0\textwidth, height=5cm]{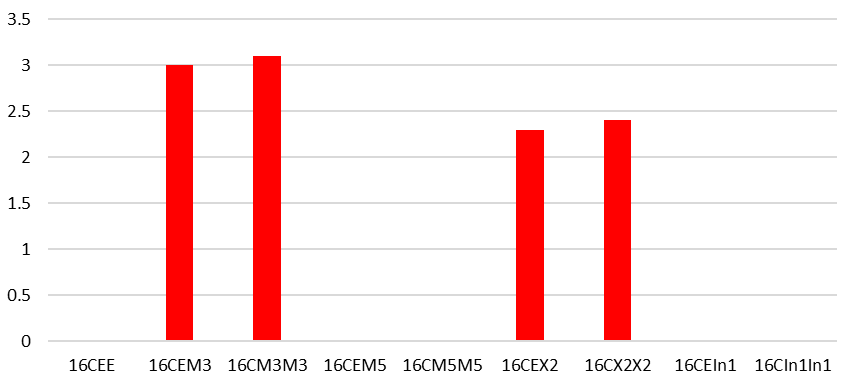}
\caption{NMED x$10^{-3}$ of 16x16 Tree Multiplier}
\label{fig:NMED_16x16Tree}
\end{figure}

\begin{figure}[!h]
\centering
\includegraphics[width=1.0\textwidth, height=5cm]{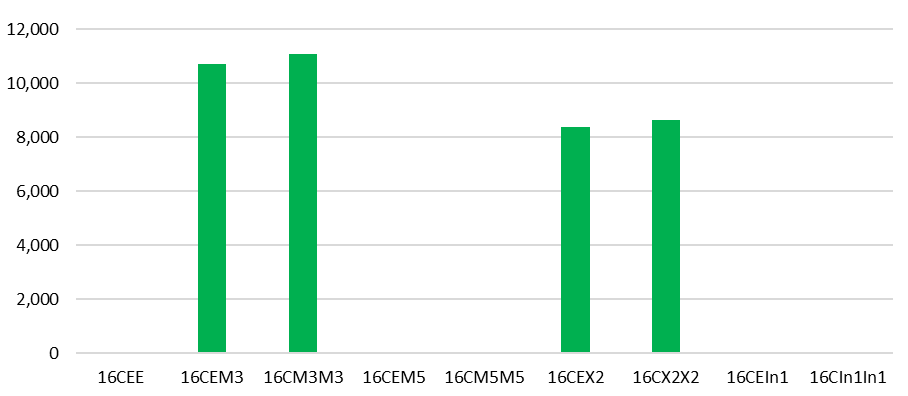}
\caption{MRED x$10^{-1}$of 16x16 Tree Multiplier}
\label{fig:MRED_16x16Tree}
\end{figure}

\begin{figure}[!h]
\centering
\includegraphics[width=1.0\textwidth, height=5cm]{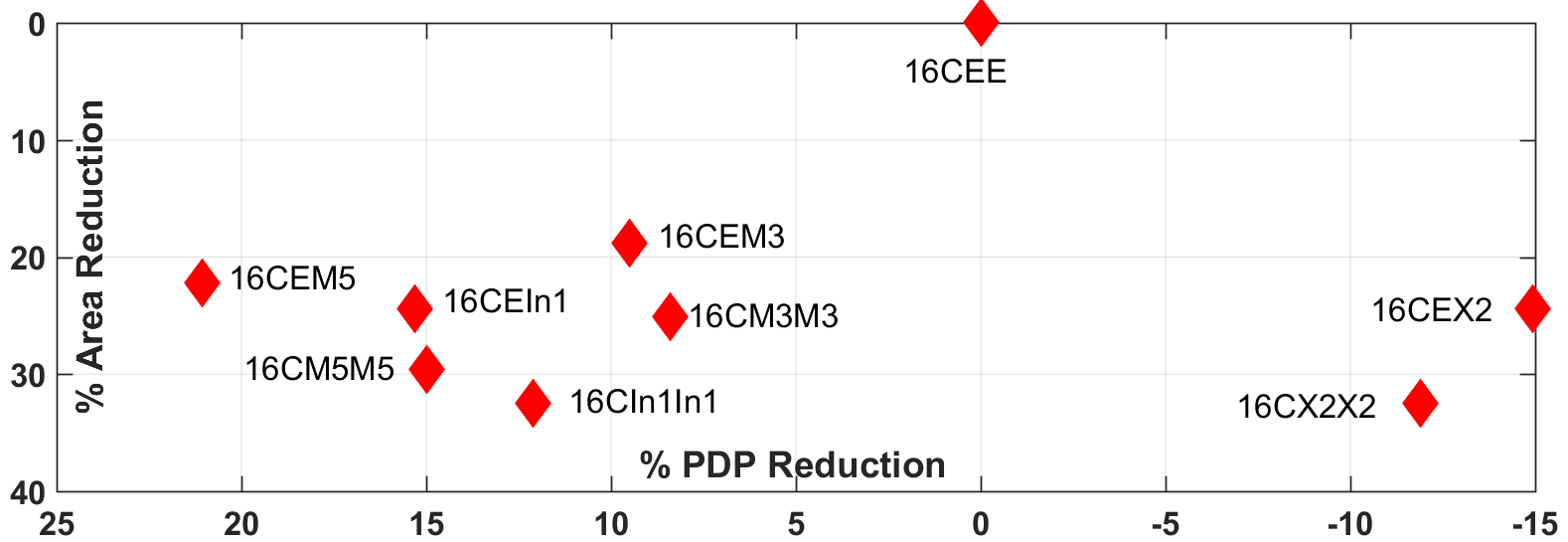}
\caption{Area and PDP Reduction of 16x16 Tree Multiplier}
\label{fig:PDP_Area_NMED_16x16Tree}
\end{figure}

Table \ref{table:16X16Tree} depicts the characterization for 16x16 approximate tree multipliers, which to some degree shows similarities to Table \ref{table:8X8Tree}. The design \textit{16CM5M5} has the lowest power consumption. Figures \ref{fig:ER_16x16Tree}, \ref{fig:NMED_16x16Tree} and  \ref{fig:MRED_16x16Tree} shows the ER, NMED and MRED for 16x16 tree multipliers, respectively. \textit{16CEIn1} and \textit{16CIn1In1} have the lowest ER, delay and area. The same designs have high NMED and MRED. As depicted in Figure \ref{fig:PDP_Area_NMED_16x16Tree} which shows area and PDP reduction, the designs on the lower left corner are superior, i.e., \textit{16CEM5}, \textit{16CEIn1} and \textit{16CM5M5} are all Pareto-designs while \textit{16CEM3} is a non-Pareto design.

\begin{table}[t!]
 \centering
 \caption{16x16 Approximate Tree Multiplier}
   \scalebox{0.7}{ 
    \begin{tabular}{|c|c|c|c|c|c|c|c|}
    \toprule
    \textbf{Type} & \multicolumn{1}{p{4.055em}|}{\textbf{MRED}} & \multicolumn{1}{p{4.055em}|}{\textbf{MED}} & \multicolumn{1}{p{4.055em}|}{\textbf{ER}} & \multicolumn{1}{p{4.055em}|}{\textbf{NMED}} & \multicolumn{1}{p{2.555em}|}{\textbf{Delay (ps)}} & \multicolumn{1}{p{2.78em}|}{\textbf{Power ($\mu$W)}} & \multicolumn{1}{p{2.11em}|}{\textbf{size}} \\
    \midrule
    \textbf{16CEE} & 00 & 00 & 00 & 00 & 680   & 100.8 & 4872 \\
    \midrule
    \textbf{16CEM3} & 1.07E+03 & 4.64E+04 & 1.00E+00 & 3.00E-03 & 663   & 93.57 & 3954 \\
    \midrule
    \textbf{16CM3M3} & 1.11E+03 & 4.80E+04 & 1.00E+00 & 3.10E-03 & 693   & 90.6  & 3648 \\
    \midrule
    \textbf{16CEM5} & 9.10E-03 & 3.90E-01 & 9.41E-01 & 2.52E-08 & 585   & 92.48 & 3792 \\
    \midrule
    \textbf{16CM5M5} & 9.30E-03 & 3.96E-01 & 9.79E-01 & 2.56E-08 & 670   & 86.98 & 3432 \\
    \midrule
    \textbf{16CEX2} & 8.37E+02 & 3.56E+04 & 1.00E+00 & 2.30E-03 & 685   & 115   & 3684 \\
    \midrule
    \textbf{16CX2X2} & 8.65E+02 & 3.71E+04 & 1.00E+00 & 2.40E-03 & 671   & 114.3 & 3288 \\
    \midrule
    \textbf{16CEIn1} & 1.74E-02 & 7.47E-01 & 8.22E-01 & 4.83E-08 & 516   & 112.5 & 3684 \\
    \midrule
    \textbf{16CIn1In1} & 1.79E-02 & 7.70E-01 & 9.04E-01 & 4.98E-08 & 527   & 114.3 & 3288 \\
    \bottomrule

 \end{tabular} 
  }
\label{table:16X16Tree}
\end{table}

\subsection{\textbf{Discussion and Comparison}}  \label{sec:Dis}
The considered approximate multipliers are implemented using Cadence's Spectre based on TSMC65nm process, with V\textsubscript{dd} = 1.0V at T=27C\textsuperscript{o}. The circuit inputs are provided by independent voltage sources, and a load of 10fF is utilized. We evaluated and compared the design characteristics (Area, Power and Delay). As shown in Tables \ref{table:8X8Array} and \ref{table:8X8Tree}, the 8x8 exact tree multiplier exhibits lower delay, power and size compared to the exact 8x8 array multiplier. 



Several multiplier designs, based on \textit{AMA5}, have the lowest delay and power consumption, due to the basic structure of the FA cell, which is composed of two buffers only. Also, they have the lowest NMED and a small size. Regarding accuracy, the designs based on \textit{InXA1} have low ER and NMED. Similarly, the designs based on the 6 transistors FA, have the minimal size. Thus, it can be observed that the characteristics of approximate FA are generally propagated in the corresponding approximate multipliers as well.




In terms of architecture, we found that the tree multiplier designs tend to have a lower power consumption than array multipliers, especially the designs based on low power consumption FAs, such as \textit{AMA3} and \textit{AMA5}. In terms of the 8x8 sub-module placement to form higher-order multipliers, with a fixed configuration for AHxBH, AHxBL and ALxBH sub-module, we have noticed that ER and NMED increase, while the size, power consumption and delay decrease for designs with a high degree of approximation in ALxBL.

Compared to the 24 different designs reported in \cite{Jiang}, where 92\% of the designs have ER close to 100\%, only 80\% of our proposed designs have high ER. Regarding NMED, almost all our designs have a value less than 10\textsuperscript{-5}, which is the minimum value reported by the 24 approximate designs in \cite{Jiang}. Comparing the PDP reduction, most of the designs in \cite{Jiang} have a high PDP reduction because they are based on truncation and a high degree of approximation. However, our designs are superior in PDP reduction for designs with a high degree of approximation.
\section{\textbf{Application}} \label{sec:application}

\begin{figure}[!t]
\centering
\includegraphics[width=14cm, height=5cm]{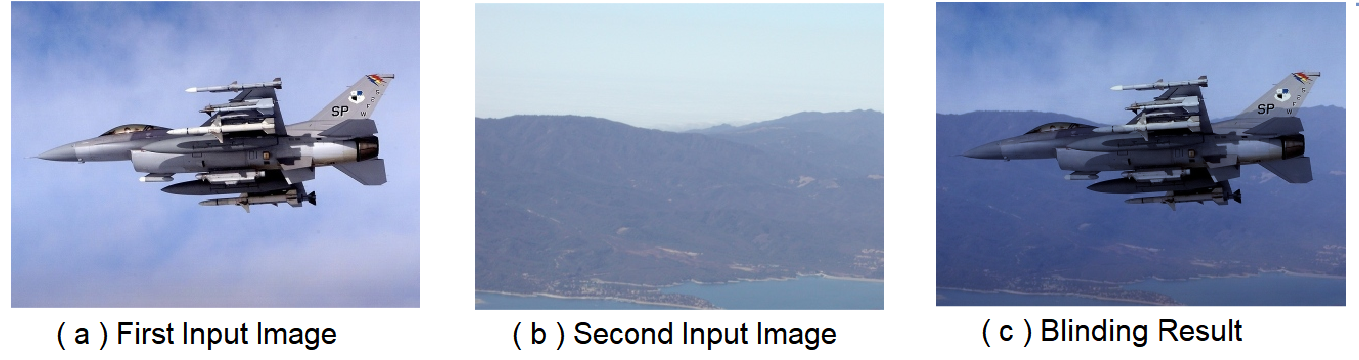}
\caption{Image blinding}
\label{fig:blinding}
\end{figure}

While in previous sections, we used Cadence Spectre to build the circuits and evaluate their area, performance and power consumption. In this section, for experimentation purposes, we evaluate and compare the accuracy of the built approximate multipliers based on an image blending application, where two images are multiplied pixel-by-pixel as shown in Figure \ref{fig:blinding}. Here, we use MATLAB to evaluate error metrics for image processing. To this end, we have modeled the same approximate multiplier circuit architectures in MATLAB and run exhaustive simulation.

\begin{figure}[h]
\centering
\includegraphics[width=1.0\textwidth, height=6cm]{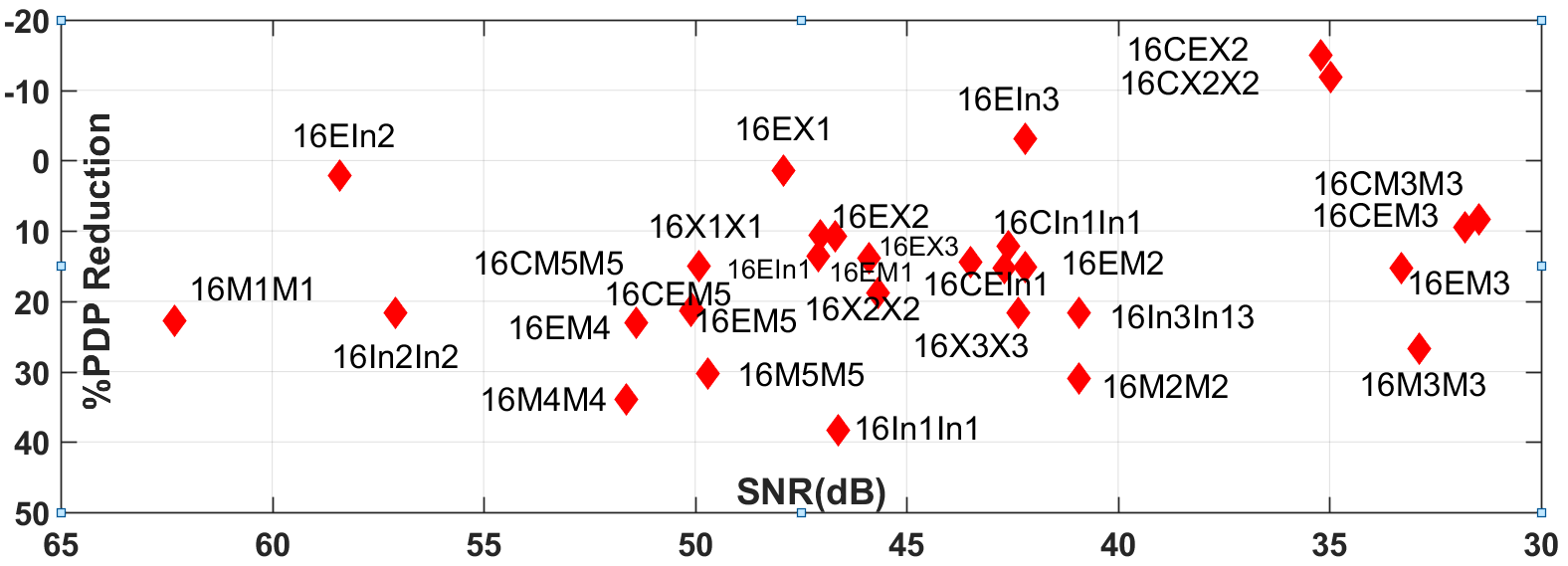}
\caption{\%PDP Reduction and SNR of Multipliers}
\label{fig:PDP_SNR}
\end{figure}

The signal to noise ratio (SNR) is used to measure the image quality. Figure \ref{fig:PDP_SNR} shows a comparison of the SNR and the percentage of PDP reduction for different approximate multipliers. Designs on the bottom left corner, have the highest PDP reduction and the best quality (high SNR). Generally, all multiplier designs have an acceptable SNR (acceptable quality). However, there exist some designs, e.g., \textit{16EIn3}, \textit{16CEX2} and \textit{16CX2X2}, with PDP greater than the exact design. The library of implemented cells and multiplier circuits, and the results of the image blending application can be found at
\textit{https://sourceforge.net/projects/approximatemultiplier}.

\section{\textbf{Conclusions}} \label{sec:conclusion}

In this report, we designed, evaluated and compared different approximate multipliers, based on approximation in partial product summation. The design space of approximate multipliers is found to be primarily dependent on the type of the approximate FA used, the architecture, and the placement of 8x8 sub-modules in the higher-order \textit{n}x\textit{n} multipliers. The proposed designs are compared based on PDP, area, delay, power, quality( ER, NMED and MRED). Various optimal designs have been identified in terms of the considered  design metrics. An image blending application is used to compare the proposed multiplier designs in terms of SNR and PDP. Our designs show comparative results compared to 24 different approximate designs reported in \cite{Jiang}. In the future, we plan to investigate the design space of higher-order multiplier modules (e.g., 64x64) using the already considered metrics and configurations. Moreover, we also plan to evaluate the possibility of having mixed FAs in the 8x8 multiplier block.

\balance
\newpage
\bibliography{mabiblio}{}

\begin{thebibliography}{10}
\providecommand{\url}[1]{#1}
\csname url@samestyle\endcsname
\providecommand{\newblock}{\relax}
\providecommand{\bibinfo}[2]{#2}
\providecommand{\BIBentrySTDinterwordspacing}{\spaceskip=0pt\relax}
\providecommand{\BIBentryALTinterwordstretchfactor}{4}
\providecommand{\BIBentryALTinterwordspacing}{\spaceskip=\fontdimen2\font plus
\BIBentryALTinterwordstretchfactor\fontdimen3\font minus
  \fontdimen4\font\relax}
\providecommand{\BIBforeignlanguage}[2]{{%
\expandafter\ifx\csname l@#1\endcsname\relax
\typeout{** WARNING: IEEEtran.bst: No hyphenation pattern has been}%
\typeout{** loaded for the language `#1'. Using the pattern for}%
\typeout{** the default language instead.}%
\else
\language=\csname l@#1\endcsname
\fi
#2}}
\providecommand{\BIBdecl}{\relax}
\BIBdecl

\bibitem{AC1}
J.~Han and M.~Orshansky, ``Approximate computing: An emerging paradigm for
  energy-efficient design,'' in \emph{European Test Symposium}, 2013, pp. 1--6.

\bibitem{M5}
P.~Kulkarni, P.~Gupta, and M.~Ercegovac, ``Trading accuracy for power with an
  underdesigned multiplier architecture,'' in \emph{VLSI Design}, 2011, pp.
  346--351.

\bibitem{speculative}
A.~K. Verma, P.~Brisk, and P.~Ienne, ``Variable latency speculative addition: A
  new paradigm for arithmetic circuit design,'' in \emph{Design, Automation
  Test in Europe}, 2008, pp. 1250--1255.

\bibitem{segmented}
N.~Zhu, W.~L. Goh, and K.~S. Yeo, ``An enhanced low-power high-sspeed adder for
  error-tolerant application,'' in \emph{Integrated Circuits}, 2009, pp.
  69--72.

\bibitem{CSA}
K.~Du, P.~Varman, and K.~Mohanram, ``High performance reliable variable latency
  carry select addition,'' in \emph{Design, Automation Test in Europe}, 2012,
  pp. 1257--1262.

\bibitem{AC2}
H.~Jiang, J.~Han, and F.~Lombardi, ``A comparative review and evaluation of
  approximate adders,'' in \emph{Great Lakes Symposium on VLSI}.\hskip 1em plus
  0.5em minus 0.4em\relax ACM, 2015, pp. 343--348.

\bibitem{Vaibhav}
V.~Gupta, D.~Mohapatra, A.~Raghunathan, and K.~Roy, ``Low-power digital signal
  processing using approximate adders,'' \emph{IEEE Transactions on
  Computer-Aided Design of Integrated Circuits and Systems}, vol.~32, no.~1,
  pp. 124--137, 2013.

\bibitem{XORFA}
Z.~Yang, A.~Jain, J.~Liang, J.~Han, and F.~Lombardi, ``Approximate
  xor/xnor-based adders for inexact computing,'' in \emph{Nanotechnology},
  2013, pp. 690--693.

\bibitem{InXA}
H.~A.~F. Almurib, T.~N. Kumar, and F.~Lombardi, ``Inexact designs for
  approximate low power addition by cell replacement,'' in \emph{Design,
  Automation Test in Europe}, 2016, pp. 660--665.

\bibitem{M6}
K.~Y. Kyaw, W.~L. Goh, and K.~S. Yeo, ``Low-power high-speed multiplier for
  error-tolerant application,'' in \emph{Electron Devices and Solid-State
  Circuits}, 2010, pp. 1--4.

\bibitem{M1}
G.~Zervakis, K.~Tsoumanis, S.~Xydis, D.~Soudris, and K.~Pekmestzi,
  ``Design-efficient approximate multiplication circuits through partial
  product perforation,'' vol.~24, no.~10, 2016, pp. 3105--3117.

\bibitem{Reddy}
K.~M. Reddy, Y.~B.~N. Kumar, D.~Sharma, and M.~H. Vasantha, ``Low power, high
  speed error tolerant multiplier using approximate adders,'' in \emph{VLSI
  Design and Test}, 2015, pp. 1--6.

\bibitem{M2}
A.~Momeni, J.~Han, P.~Montuschi, and F.~Lombardi, ``Design and analysis of
  approximate compressors for multiplication,'' in \emph{IEEE Transactions on
  Computers}, vol.~64, no.~4, 2015, pp. 984--994.

\bibitem{Jiang}
H.~Jiang, C.~Liu, N.~Maheshwari, F.~Lombardi, and J.~Han, ``A comparative
  evaluation of approximate multipliers,'' in \emph{Nanoscale Architectures},
  2016, pp. 191--196.

\bibitem{Mrazek}
R.~Hrbacek, V.~Mrazek, and Z.~Vasicek, ``Automatic design of approximate
  circuits by means of multi-objective evolutionary algorithms,'' in
  \emph{Design and Technology of Integrated Systems}, 2016, pp. 1--6.

\bibitem{Pareto}
Z.~Yang, J.~Yang, K.~Xing, and G.~Yang, ``Approximate compressor based
  multiplier design methodology for error-resilient digital signal
  processing,'' in \emph{2016 IEEE International Conference on Signal and Image
  Processing (ICSIP)}, 2016, pp. 740--744.

\bibitem{Rehman}
S.~Rehman, W.~El-Harouni, M.~Shafique, A.~Kumar, and J.~Henkel,
  ``Architectural-space exploration of approximate multipliers,'' in
  \emph{CAD}.\hskip 1em plus 0.5em minus 0.4em\relax ACM, 2016, pp. 1--8.

\bibitem{rabaey}
J.~M. Rabaey, A.~Chandrakasan, and B.~Nikolic, \emph{Digital Integrated
  Circuits}.\hskip 1em plus 0.5em minus 0.4em\relax Prentice-Hall, 2002.

\bibitem{Shao}
B.~Shao and P.~Li, ``Array-based approximate arithmetic computing: A general
  model and applications to multiplier and squarer design,'' vol.~62, no.~4,
  2015, pp. 1081--1090.

\bibitem{SALSA}
S.~Venkataramani, A.~Sabne, V.~Kozhikkottu, K.~Roy, and A.~Raghunathan,
  ``{SALSA}: Systematic logic synthesis of approximate circuits,'' in
  \emph{Design Automation Conference}, 2012, pp. 796--801.

\bibitem{SASIMI}
S.~Venkataramani, K.~Roy, and A.~Raghunathan, ``Substitute-and-simplify: A
  unified design paradigm for approximate and quality configurable circuits,''
  in \emph{Design, Automation Test in Europe}, 2013, pp. 1367--1372.

\bibitem{Parhami}
B.~Parhami, \emph{Computer Arithmetic: Algorithms and Hardware Designs}.\hskip
  1em plus 0.5em minus 0.4em\relax Oxford University Press, 2010.

\bibitem{FEMTO}
D.~Sengupta and S.~S. Sapatnekar, ``{FEMTO}: Fast error analysis in multipliers
  through topological traversal,'' in \emph{International Conference on
  Computer-Aided Design}, 2015, pp. 294--299.

\end{thebibliography}
\bibliographystyle{IEEEtran}
\end{document}